\documentclass[a4paper,fleqn,usenatbib]{mnras}

\usepackage[T1]{fontenc}
\usepackage{ae,aecompl}

\usepackage{graphicx}	
\usepackage{amsmath}	
\usepackage{amssymb}	

\newcommand{\eff}{_{\textrm{\tiny eff}}} 
\newcommand{\subsun}{_{\sun}} 
\usepackage{hyperref}

\title[The sdA problem -- II. Observational Follow-up]{The sdA problem -- II. Photometric and Spectroscopic Follow-up}

\author[I. Pelisoli et al.]{
	Ingrid Pelisoli$^{1}$\thanks{E-mail: ingrid.pelisoli@ufrgs.br},
	S. O. Kepler$^{1}$,
	D. Koester$^{2}$,
	B. G. Castanheira$^{3,4}$,
	\newauthor A. D. Romero$^{1}$,
	L. Fraga$^{5}$
	\\
	$^{1}$Instituto de F\'{i}sica, Universidade Federal do Rio Grande do Sul, 91501-900, Porto-Alegre, RS, Brazil\\
	$^{2}$Institut f\"{u}r Theoretische Physik und Astrophysik, Universit\"{a}t Kiel, D-24098, Kiel, Germany\\
	$^{3}$Baylor University, Waco, TX - 76798, USA\\
    $^{4}$Department  of  Astronomy,  University  of  Texas  at  Austin, Austin, TX - 78712, USA\\
	$^{5}$Laborat\'{o}rio Nacional de Astrof\'{i}sica LNA/MCTIC, 37504-364, Itajub\'{a}, MG, Brazil
}

\date{Accepted XXX. Received YYY; in original form ZZZ}

\pubyear{2018}

\begin{document}
\label{firstpage}
\pagerange{\pageref{firstpage}--\pageref{lastpage}}
\maketitle

\begin{abstract}
Subdwarf A star (sdA) is a spectral classification given to objects showing H-rich spectra and sub-main sequence surface gravities, but effective temperature lower than the zero-age horizontal branch. Their evolutionary origin is an enigma. In this work, we discuss the results of follow-up observations of selected sdAs. We obtained time resolved spectroscopy for 24 objects, and time-series photometry for another 19 objects. For two targets, we report both spectroscopy and photometry observations. We confirm seven objects to be new extremely-low mass white dwarfs (ELMs), one of which is a known eclipsing star. We also find the eighth member of the pulsating ELM class.
\end{abstract}




\section{Introduction}

White dwarf stars are the most common outcome of single-star evolution, corresponding to the final observable evolutionary stage of all stars with initial mass below $7-10.6$~$M\subsun${} \citep[e.g.][]{woosley2015}, including the Sun and over 95~per cent of all stars in the Galaxy. Their relative abundance, combined with a simple structure and long cooling timescales, makes them the perfect laboratory for modelling stellar evolution \citep[e.g.][]{kalirai2008,romero2015}, and for population synthesis studies constraining the age and star formation history of different stellar populations \citep[e.g.][]{liebert2005,tremblay2016,kilic2017}. About 25 per cent of white dwarfs in the galactic field are known to have a companion \citep{toonen2017}, therefore white dwarfs have also the potential to put constraints into binary evolution channels.

Short-period binary white dwarfs in particular are potential progenitors of Type Ia \citep{webbink1984,iben1984} and .Ia supernovae \citep{bildsten2007}. This fact motivated the first surveys for white dwarfs in close binaries \citep{robinson1987,foss1991}, which resulted on null detections. The first successfull survey was performed by \citet{marsh1995}. They noticed that the catalogue of \citet{bergeron1992} contained fourteen white dwarfs with spectroscopic mass below $0.45$~$M\subsun$, which cannot be formed within a Hubble time without some form of mass-loss enhancement. They were most likely the remnants of mass transfer in post-main-sequence common-envelope binaries. Indeed, \citet{marsh1995} confirmed five out of the seven stars they probed to be in binaries. More recent studies suggest that the binary fraction of low-mass white dwarfs ($M \lesssim 0.45$~$M\subsun$) is at least 70 per cent \citep{brown2011}. Low-mass single systems can be explained by other mass-loss enhancing mechanisms, such as high metallicity \citep{dcruz1996} or supernova stripping \citep{wang2009}, by mass ejection caused by a massive planet \citep{nelemans1998}, or by merger events \citep{zhang2012,zhang2017}. For the currently known white dwarfs with mass below 0.3~$M\subsun$, the binary fraction seems to be close to 100~per cent \citep{elmsurveyVII}. These systems are known as extremely-low mass white dwarfs (ELMs).

The ELM Survey \citep{elmsurveyI, elmsurveyII, elmsurveyIII, elmsurveyIV, elmsurveyV, elmsurveyVI, elmsurveyVII} made large progress in the study of these objects. They have found 88 systems, 76 of which were confirmed to be in binaries, mostly through the analysis of their radial velocity (RV) variations. Seven systems were found to be pulsators, eight show ellipsoidal variations, and two are eclipsing systems \citep{hermes2012, hermes2013a, hermes2013b, kilic2015, bell2015, elmsurveyVII}. The obtained distribution of secondary mass suggests that over 95~per cent of the systems are not Type Ia supernovae progenitors \citep{elmsurveyVII}. They are, nonetheless, strong gravitational wave sources \citep{elmsurveyIV}, given that most systems will merge within a Hubble time \citep{brown2016}. The gravitational wave radiation of the shortest orbital period systems ($P \lesssim 1$~h) may be directly detected by upcoming space-based missions such as the Laser Interferometer Space Antenna (LISA). \citet{elmsurveyIV} found three systems that should clearly be detected by missions like LISA in the first year of operations. Three other systems are above the proposed 1-$\sigma$ detection limit after one year of observations. Even when not significantly above the detection limit, ELMs are important indicators of what the Galactic foreground may look like for these detectors. Therefore understanding the space density, period distribution, and merger rate of these systems is crucial for interpreting the results of upcoming space-based gravitational wave missions, and for studying the evolution of interacting binary systems.

\begin{figure*}
	\includegraphics[angle=-90,width=\textwidth]{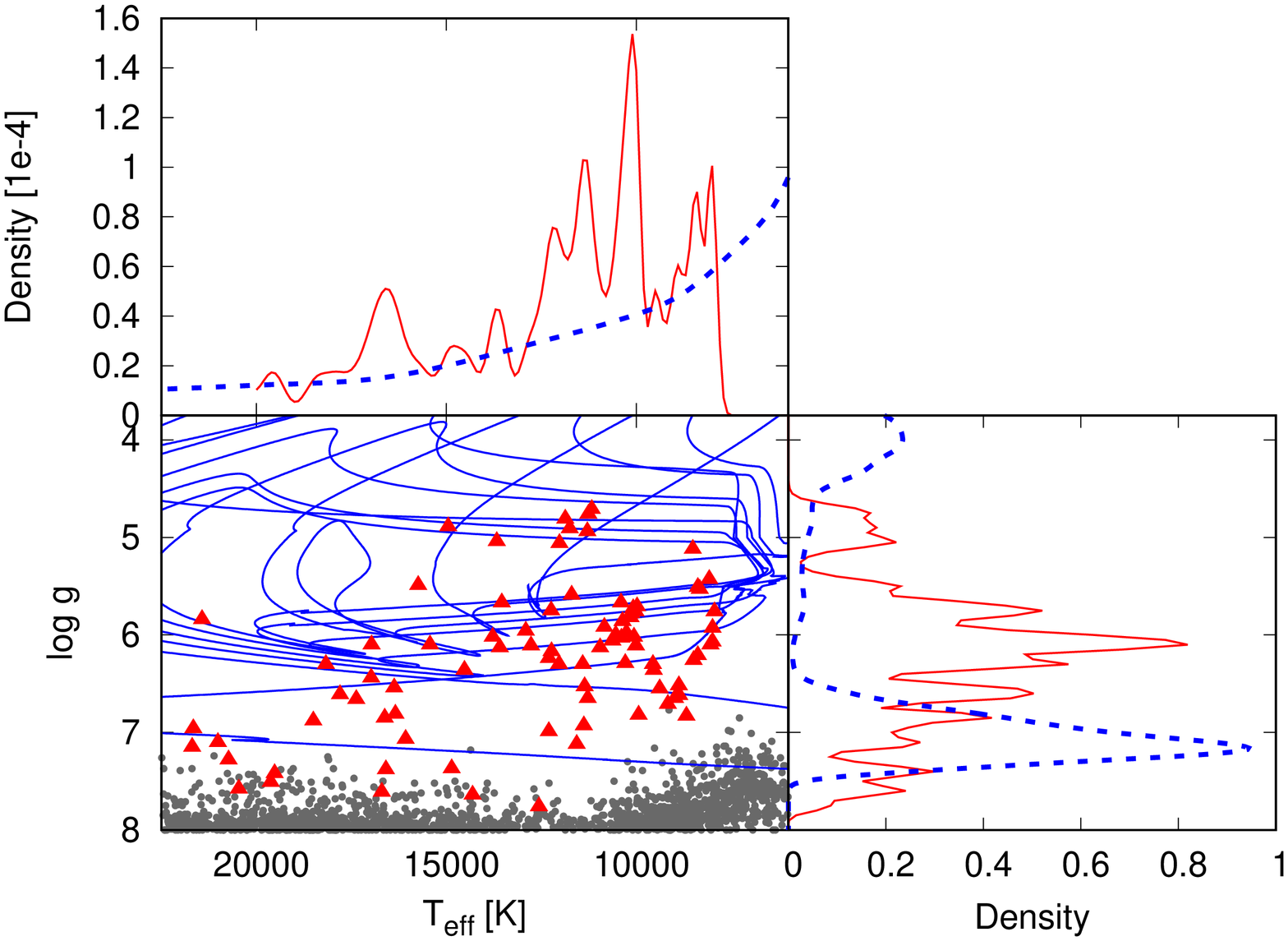}
	\caption{Bottom-left panel shows the $T\eff - \log~g$ diagram for the objects in the ELM Survey, shown as red triangles, compared to two binary evolution models of \citet{istrate2016} (blue lines), resulting on ELMs of masses 0.182 and 0.324~$M\subsun$. The white dwarfs from \citet{dr12cat} are shown as grey dots for comparison. The top panel shows the distributions in $T\eff$, both for the observed ELMs (red continuous line) and obtained from the models (blue dashed line). The bottom-right panel shows the distributions for $\log~g$. The distributions for the models were obtained taking into account the time spent at each bin of $T\eff$ or $\log~g$ compared to the total evolutionary time, with a spherical volume correction to account for difference in brightness \citep[see][for further details]{pelisoli2018}. Note that there is a lack of known ELMs in the low $T\eff$ and low~$\log g$ ends of the distribution. There are also missing objects around $\log~g \sim 7.0$; however, this range can also be reached through single evolution.}    
	\label{fig1}
\end{figure*}

The target selection of the ELM Survey was initially developed to find B-type hypervelocity stars \citep[see the MMT Hypervelocity Star Survey:][]{HVSurveyI,HVSurveyII,HVSurveyIII}, hence it favours the detection of hot ELMs ($T\eff \gtrsim 12\,000$~K). Cooler objects ($T\eff \lesssim 10\,000$~K) were targeted by \citet{elmsurveyIII}. Yet, less than 5~per cent of the objects in the ELM Survey show $T\eff \lesssim 9\,000$~K, while evolutionary models \citep{althaus2013,corsico2014,corsico2016,istrate2016} predict the same amount of time to be spent above and below said $T\eff$. Although the uncertainties in residual burning can significantly influence the cooling timescale, it is still expected that 20--50~per cent of the ELMs should show $T\eff < 9\,000$~K \citep{brown2017,pelisoli2017}. Moreover, the ELM Survey selection criteria also favoured higher $\log g$ objects. The low-$\log~g$ phases happen before the object reaches the white dwarf cooling track \citep[the objects are hence known as pre-ELMs, see e.g.][]{maxted2011,maxted2014} and are relatively quick. However, pre-ELMs are also much brighter. Assuming a spherical distribution, we found in \citet{pelisoli2017} that there should be about a hundred detected objects with $\log~g = 5-6$ for each object with $\log~g = 6-7$ in a magnitude-limited survey. Hence there is clearly a missing population of cool, low-mass ELMs yet to be found, as evidenced in Fig. \ref{fig1}.

In an effort to retrieve these missing objects, \citet{dr12cat} extended their white dwarf catalogue down to $\log g = 5.5$, revealing a population of objects which were dubbed subdwarf A stars (sdAs). Their spectra are dominated by hydrogen lines, suggestive of $T\eff \sim 10\,000$~K and of $4.75 < \log g < 6.5$. \citet{brown2017} suggested they are mainly metal-poor A/F stars in the halo with an overestimated $\log g$, given the pure hydrogen grid used to fit these objects in \citet{dr12cat}. However, as we showed in \citet{pelisoli2018}, the addition of metals to the models does not necessarily lower the estimated $\log g$. Moreover, we identified the existence of clearly two population within the sdAs, with overlapping but distinct colour distributions, and found that at least 7~per cent of the sdAs are more likely (pre-)ELMs than main sequence stars, given their physical and kinematic parameters. The missing (pre-)ELMs are thus likely within the sdA population.

In this work, we follow up on selected sdAs to probe their binarity with the aim of extending the population of known (pre-)ELMs to all the space of physical parameters predicted by the evolutionary models. We obtain both time-resolved spectroscopy to search for radial velocity (RV) variations indicating the presence of a close binary companion, and time-series photometry to look for eclipses, ellipsoidal variations, or pulsations typical of ELMs. Extending the sample of known ELMs to cool temperatures and lower masses will allow us to more robustly test the evolutionary models. With a more complete sample, we will also be able to make more reliable predictions as to the contribution of the gravitational wave signals from ELMs to upcoming missions.

\section{Methods}

\subsection{Observations}

Our observing campaign targeted bright objects with $T\eff$ and $\log g$ in the range predicted by the evolutionary models shown in Fig. \ref{fig1}. Targets with high proper motion and/or high radial velocities, and ELM-like colours, yielding high probability of being a (pre-)ELM according to \citet{pelisoli2018}, were prioritized. We have used the proper motions from the GPS1 catalogue \citep{tian2017}, which contains all but one of the objects analysed here. Consistency checks were done with the Hot Stuff for One Year \citep[HSOY,][]{altmann2017} and the UCAC5 \citep{zacharias2017} catalogues. We found the proper motions to agree within uncertainties for all the objects studied here. Priority was also given to objects showing radial velocity variations in the subspectra taken by the Sloan Digital Sky Survey (SDSS). We obtained time-resolved spectroscopy for 26 targets. We have also obtained time-series photometry for 21 targets in the vicinity of the instability strip by \citet{tremblay2015} and \citet{elmsurveyVI}, which was empirically obtained taking into account 3D corrections to $T\eff$ and $\log~g$. The targets are listed in Table~\ref{kinematic}, as well as their SDSS $g$ magnitde, proper motion (ppm), distances given a MS or (pre-)ELM radius and velocities in the Galactic rest frame ($v_{los}$).

\begin{table*}
	\centering
	\caption{Followed-up objects and some notable kinematic properties, as well as the SDSS $g$ magnitude. The first 24 objects were followed-up spectroscopically only, next 19 only photometrically, and the last two both photometrically and spectroscopically. The quoted proper motions are from \citet{tian2017}, except for J110338.46-160617.4, which is not in this catalogue and whose proper motion was then obtained from \citet{altmann2017}. Many objects show unreliable proper motion given the relatively faint ($g \gtrsim 18$) magnitude. Note that most objects would be tens of kpc away given a MS radius.}
	\label{kinematic}
	\begin{tabular}{ccccccc}
		\hline
		\hline
		SDSS~J & $g$ & ppm (mas/yr) & $\sigma_{\textrm{ppm}}$ (mas/yr) & $d_{\textrm{MS}}$ (pc) & $d_{\textrm{(pre-)ELM}}$ (pc) & $v_{los}$ (km/s) \\
		\hline
		004227.73-010634.9 & 18.63 & 1.1 & 2.5 & 16494 & 254 & 111 \\ 
		011508.65+005346.1 & 18.07 & 7.5 & 1.6 & 16799 & 232 & -173 \\ 
		024932.84-010708.4 & 19.44 & 3.2 & 2.5 & 27852 & 405 & -146 \\ 
		030608.92-001338.9 & 16.95 & 19.3 & 2.0 & 6625 & 108 & 189 \\ 
		032914.77+003321.8 & 16.76 & 11.5 & 2.3 & 21652 & 202 & 88 \\ 
		045515.00-043231.0 & 16.49 & 8.8 & 2.3 & 8090 & 113 & 64 \\ 
		073934.37+172225.5 & 18.07 & 1.4 & 2.2 & 10429 & 173 & -65 \\ 
		084034.83+045357.6 & 17.34 &  2.7 & 1.6 & 9004 & 140 & 9 \\ 
		090410.00+034332.9 & 17.58 &  3.3 & 3.1 & 9697 & 153 & -128 \\ 
		092056.09+013114.8 & 16.53 & 12.0 & 2.1 & 5346 & 87 & -125 \\ 
		101701.89+070806.8 & 18.25 & 6.2 & 2.9 & 17876 & 250 & 253 \\ 
		112616.66-010140.7 & 18.50 & 5.5 & 2.0 & 17390 & 256 & 62 \\ 
		112620.47+090145.5 & 18.85 & 6.2 & 2.5 & 23810 & 331 & 253 \\ 
		122911.49-003814.4 & 18.27 & 10.2 & 1.8 & 14079 & 218 & 380 \\ 
		142421.30-021425.4 & 16.93  & 22.5 & 1.8 & 12867 & 159 & 14 \\ 
		155937.48+113721.9 & 17.22 & 7.4 & 1.8 & 30696 & 270 & 31 \\ 
		162624.91+162201.5 & 17.04 & 6.8 & 1.8 & 27991 & 247 & -27 \\ 
		205120.67+014554.4 & 17.27 & 9.0 & 2.4 & 8954 & 138 & 110 \\ 
		213428.63-011409.3 & 16.96 & 3.9 & 2.3 & 26771 & 237 & 142 \\ 
		223831.91+125318.3 & 15.55 & 13.6 & 1.6 & 3946 & 61 & -18 \\ 
		233343.95-001502.0 & 19.32 & 2.2 & 2.8 & 26381 & 382 & -16 \\ 
		233403.21+153829.2 & 16.34 & 39.1 & 1.5 & 3435 & 62 & 12 \\ 
		233606.13-102551.5 & 19.34 & 5.1 & 2.7 & 25693 & 378 & 95 \\ 
		233708.62-094307.0 & 17.90 & 4.1 & 1.7 & 12642 & 191 & -157 \\ 
		\hline
		045001.34-042712.9 & 19.07 & 8.0 & 3.4 & 19701 & 308 & -38 \\ 
		073958.57+175834.4 & 14.75 & 8.7 & 2.5 & 69162 & 236 & -49 \\ 
		075133.48+101809.4 & 17.40 & 2.1 & 2.1 & 37588 & 314 & -46 \\ 
		075519.92+091511.0 & 15.32 & 3.5 & 1.7 & 2729 & 46 & -131 \\ 
		075738.94+144827.5 & 15.04 & 1.1 & 1.4 & 3415 & 51 & -84 \\ 
		092140.37+004347.9 & 18.39 & 1.7 & 2.1 & 13550 & 215 & -64 \\ 
		094144.89+001233.8 & 19.28 & 8.1 & 2.6 & 22149 & 345 & -50 \\ 
		104522.80-023735.6 & 19.28 & 7.2 & 2.8 & 15689 & 275 & 101 \\ 
		110338.46-160617.4 & 15.77 & 7.3 & 3.3 & 5154 & 75 & 184 \\ 
		111041.50+132354.3 & 18.28 & 19.0 & 2.4 & 9648 & 170 & 262 \\ 
		112058.97+042012.3 & 17.87 & 1.2 & 2.1 & 51622 & 412 & 158 \\ 
		140353.33+164208.1 & 16.20 & 7.7 & 1.5 & 5748 & 85 & 16 \\ 
		143333.45+041000.8 & 18.31 & 2.2 & 2.3 & 21349 & 281 & 182 \\ 
		160040.95+102511.7 & 15.00 & 7.1 & 1.2 & 3092 & 47 & 94 \\ 
		163625.08+113312.4 & 17.24 & 13.0 & 1.0 & 12043 & 164 & -215 \\ 
		165700.89+130759.6 & 15.62 & 1.5 & 1.4 & 4671 & 68 & 63 \\ 
		201757.29-125615.6 & 17.07 & 4.8 & 1.7 & 8796 & 131 & 168 \\ 
		204038.41-010215.7 & 16.59 & 9.1 & 1.7 & 6381 & 98 & 165 \\ 
		233625.92+150259.6 & 17.18 & 2.6 & 1.9 & 9003 & 134 & -160 \\ 
		\hline
		134336.44+082639.4 & 16.34 & 17.5 & 2.1 & 5071 & 81 & 352 \\ 
		222009.74-092709.9 & 15.81 & 9.6 & 1.6 & 4659 & 71 & 78 \\ 
		\hline
		\hline
	\end{tabular}
\end{table*}

We carried out spectroscopy mainly with the Goodman Spectrograph \citep{goodman} on the 4.1~m Southern Astrophysical Research (SOAR) Telescope. All exposures were taken with a 1.0'' slit, and binned by a factor of two in both dimensions. We used a 1200~l/m grating, with a camera angle of 30.00$^{\circ}$ and grating angle of 16.30$^{\circ}$, obtaining a wavelength coverage of 3600--4950~\AA{} with a resolution of $\sim 2$~\AA.

We also obtained spectroscopy with the GMOS spectrographs  \citep{hook2004,gimeno2016} on both Gemini North and Gemini South 8.1~m telescopes. The exposures were taken with a 0.75" slit. As with SOAR, we binned the CCD by a factor of two in both dimensions and used a 1200~l/mm grating. Exposures centred at both 4400~\AA{} and 4450~\AA{} were taken at each semester, to dislocate the position of the two gaps between the CCDs in GMOS, covering wavelengths 3580--5190~\AA{} and 3630--5240~\AA, respectively. Our data was partially affected by the bright columns issue developed by GMOS-S CCD2 and CCD3 during September 30, 2016 -- February 21, 2017.

Five $\log g > 5.5$ objects were observed with the medium resolution echelle spectrograph X-shooter \citep{vernet2011}, mounted on VLT-UT2 at Paranal, Chile. X-shooter covers the spectral range from the atmospheric cut-off in the UV to the near-infrared with three separate arms: UVB (3000 -- 5600~\AA), VIS (5600 -- 10100~\AA) and NIR (10100 -- 24000~\AA). The data were taken in stare mode, using slits of 1.0'', 0.9'', and 1.2'' for UVB, VIS, and NIR arms, respectively, which allows a resolution of $\sim 1$~\AA.  X-shooter has the advantage of also allowing to search for red companions, that could appear as an excess in the NIR arm spectra.

For all instruments, arc-lamp exposures were taken before and after each science exposure to verify the stability. For the wavelength calibration, a CuHeAr lamp was taken after each round of exposures, at the same position of the science frames. Due to the faintness of the objects and the need for multiple spectra, the exposure time was estimated aiming at a median signal-to-noise ratio ($S/N$) of 10--15 per exposure. One radial velocity standard was observed at each semester to verify the reliability of the method, and a spectrophotometric standard star was observed every night for the flux calibration, except for Gemini observations, which observed one spectrophotometric standard star per semester.

Time-series photometry was obtained with the 1.6~m Perkin-Elmer telescope at Observat\'{o}rio do Pico dos Dias (OPD, Brazil), with an Andor iXon CCD and a red-blocking filter (BG40). We have also used the imaging mode in Goodman at SOAR for photometry, with the S8612 red-blocking filter. The integration time varied from 10 to 30~s, depending on the brighness of the target, with typical readout of 1-3~s.

\subsection{Data analysis}

SOAR spectroscopic data were reduced using {\sc iraf}'s {\sc noao} package. The frames were first bias-subtracted, and flattened with a quartz lamp flat. We then extracted the spectra and did the wavelength calibration with a CuHeAr lamp spectrum extracted with the same aperture. Finally, flux and extinction calibration were applied. The {\sc Gemini iraf} package was used for data from these telescopes, and the X-shooter pipeline for the VLT data, with equivalent steps in the reduction.

Radial velocity estimates were done with the {\sc xcsao} task from the {\sc rvsao} package \citep{kurtz1998}, after verifying that the intercalated HeAr lamps presented no shift, which was always the case. We cross-correlated the spectral region covering all visible Balmer lines (typically from 3750 to 4900~\AA) to spectral templates from the updated model grid based on \cite{koester2010}, described in \cite{pelisoli2018}. The values of RV were corrected to the Solar System barycentre given the time of observations and the telescope location. All our RV estimates are given in Table~\ref{RVs}. We have not added the RV estimated from the SDSS spectra to our dataset, because it was obtained eight years before our data even in the best of cases, hence the phase might not be accurate to our recently obtained data.

We performed a Shapiro-Wilk normality test \citep{shapirowilk} to verify whether the obtained velocities displayed a behaviour which could be explained by Gaussian uncertainties. Next, we calculated the Lomb-Scargle periodogram \citep{lomb1976,scargle1982} using the NASA Exoplanet Archive tool\footnote{\url{https://exoplanetarchive.ipac.caltech.edu/cgi-bin/Pgram/nph-pgram}}. For each of the highest fifty peaks in the periodogram, we calculated an orbital solution of the form
\begin{eqnarray}
RV (t) = RV_0 + K\sin(2\pi t/T + \phi),
\end{eqnarray}
where $RV_0$ is the systemic velocity, $K$ is the semi-amplitude of the RV variation, $T$ is the period, and $\phi$ the phase. We selected as the best solution the one with the highest reduced $R^{2}$, defined as
\begin{eqnarray}
R^2 \equiv 1 - \frac{\sum_i (y_i - f_i)^2 }{\sum_i (y_i - \bar{y})^2 },
\end{eqnarray}
where $y_i$ are the observed values, $\bar{y}$ is their mean, and $f_i$ are the adjusted values. This is equivalent to selecting the solution with the smallest reduced $\chi^2$.

Each individual spectrum was later Doppler-corrected considering the estimated velocities, and all spectra of each object were combined to obtain a $S/N \gtrsim 30$ spectrum (the average for the whole sample was $S/R = 45$). Considering the lack of strong metal lines, we fit these spectra to a grid of models assuming metallicity $Z = 0.1~$Z$\subsun$, with the same input physics as described in \citet{pelisoli2018}. We caution that quoted uncertainties are formal fitting errors, and the systematic uncertainties are larger. We previously estimated it to be $\sim5$~per cent in $T\eff$ and $ 0.25$~dex in $\log~g$ \citep[e.g.][]{pelisoli2018}; however, as we will show in Section~\ref{fits}, it seems that the systematic uncertainty in $\log~g$ can actually be higher in the $T\eff-\log~g$ region of the sdAs, and reach 0.5~dex. The SDSS spectra of all objects were also fit to the same $Z = 0.1~$Z$\subsun$ grid to allow a comparison. We have relied on the SDSS colours to choose between hot and cool solutions with similar $\chi^2$, that arise due to similar equivalent width being possible with different combinations of $T\eff$ and $\log~g$. To estimate the mass of each object, we interpolated the models of \citet{althaus2013}. The models of \citet{istrate2016} made a large improvement to the input physics, by taking into rotational mixing, which was shown to be an important factor in the atmosphere abundances for ELMs. However, the lowest ELM mass in the models of \citet{istrate2016} is $0.16-0.18~M\subsun$, depending on the metallicity, and most of our objects show mass lower than that. Only one object (SDSSJ1626+2622) could have its mass accurately determined with \citet{istrate2016} models, and it agreed with the mass estimate using \citet{althaus2013} within uncertainties. Hence, to be consistent, we used the models of \citet{althaus2013} for all mass estimates.

All photometry images were bias-subtracted, and flat-field corrected using dome flats. Aperture  photometry  was  done  using  the {\sc daophot} package in {\sc iraf}. A neighbouring non-variable star of similar brightness was used to perform differential photometry. The resulting light curve was analysed with {\sc Period}04 \citep{lenz2005}, in search of pulsations with amplitude at least four times larger than the average amplitude of the Fourier transform. {\sc Period}04 was also used to fit the light curve and perform pre-whitening when pulsations were found, and to estimate uncertainties using the Monte Carlo method with a 1000 simulations.

\section{Results}

We found seven objects whose RV variations indicate they are in close binaries. They show p-value smaller than 0.15 for the Shapiro-Wilk test, implying that the variations cannot be explained by Gaussian noise to a confidence level of 85 per cent. For six objects out of these seven, the p-value is smaller than 0.05, hence the confidence level is 95 per cent. The orbital solution shows $R^2$ larger than 0.95 for all but one object. $T\eff$ and $\log g$ suggest they are new (pre-)ELMs (see Fig.~\ref{fig2}). Their properties are given in Section~\ref{newELMs}.

For six other objects, the p-value is larger than 0.15, but we obtain an orbital solution with a short period ($P \lesssim 10$~h), expected from (pre-)ELMs in the range of physical parameters for the sdAs \citep{brown2017}, and $R^2 \gtrsim 0.85$. Two other objects shows $p<0.05$, but its atmospheric parameters are compatible with both a pre-ELM and a main sequence star. More data is required to confirm the nature of these eight objects; given the distance modules or proper motion and the estimated physical parameters, we assume they are probable (pre-)ELMs and discuss their properties in Section~\ref{probELMs}.

Six other objects have twelve measurements or more \citep[the average necessary to confirm binariy, according to][]{elmsurveyVII}, in at least three different epochs and often multiple telescopes, but the Shapiro-Wilk test suggested no real variation. These objects are possibly single stars, or show either very short ($\lesssim 1.0$~h) or long periods ($\gtrsim 200$~days). We also found no RV variation or red companions for the five objects observed with X-shooter in three nights over a week. All these objects are detailed in Section~\ref{single}.

In Section~\ref{fits}, we compare the values of $T\eff$ and $\log g$ obtained fitting the SDSS spectra and the SOAR or X-shooter spectra for each object. We were unable to obtain a good fit to the Gemini spectra. The Gemini reduction package interpolates between the CCD gaps before performing the flux calibration, and that seems to be affecting the output to a point where our models cannot fit the slope of the continuum.

In addition, we have found seven new pulsators among the sdAs. For other fourteen observed stars, we have obtained a detection limit $\lesssim 10$~mmag and found no pulsations. The photometry results are discussed in Section~\ref{phot}.

\subsection{New (pre-)ELMs}
\label{newELMs}

\subsubsection{J032914.77+003321.8} 

\begin{figure}
	\includegraphics[angle=-90,width=\columnwidth]{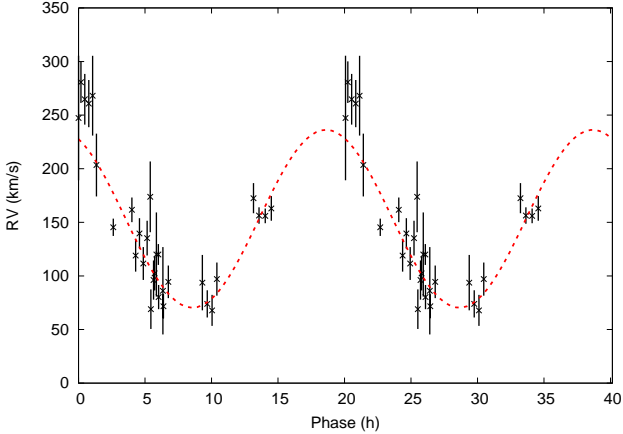}
	\caption{Orbital solution for SDSS~J032914.77+003321.8, phase-folded to the 20.1~h period. Note that in this and also in the next orbital solution plots, two cycles are shown. The semi-amplitude is $83\pm22$~km~s$^{-1}$, and the systemic velocity is  $153\pm18$~km~s$^{-1}$.}
	\label{J0329}
\end{figure}

The thirty RV estimates for J0329+0033, taken over seven non-consecutive nights at SOAR, yielded a Shapiro-Wilk $p$-value of 0.004, suggesting with a very high confidence level that the observed variations are not due to chance. We have estimated the period to be $20.1\pm0.1$~h. The semi-amplitude is not well constrained by our data; we estimated it to be $83\pm22$~km~s$^{-1}$. This results on an orbital fit with $R^2=0.78$, shown in Fig.~\ref{J0329}, the lowest $R^2$ among our fits. However, when we assume a main sequence radius, the photometric parallax gives a distance larger than 20~kpc for this object, what is inconsistency with its proper motion of $11.5\pm2.3$~mas~yr$^{-1}$\citep{tian2017}. The systemic velocity is also relatively high, $153.3\pm18$~km~s$^{-1}$. Assuming an ELM radius, the distance drops to $\sim 200$~pc.

Our fit to the SOAR spectrum of J0329+0033 gives $T\eff = 9080\pm10$~K and  $\log~g =  5.18\pm0.03$. Interpolating the models of \cite{althaus2013}, we obtain $M = 0.1536\pm0.0006~M\subsun$. Given this mass and the orbital parameters, the minimal mass of the companion (for an edge-on orbit) is $M_2 = 0.17~M\subsun$, implying a merging time shorter than 765~Gyr.


\subsubsection{J073934.37+172225.5} 

\begin{figure}
	\includegraphics[angle=-90,width=\columnwidth]{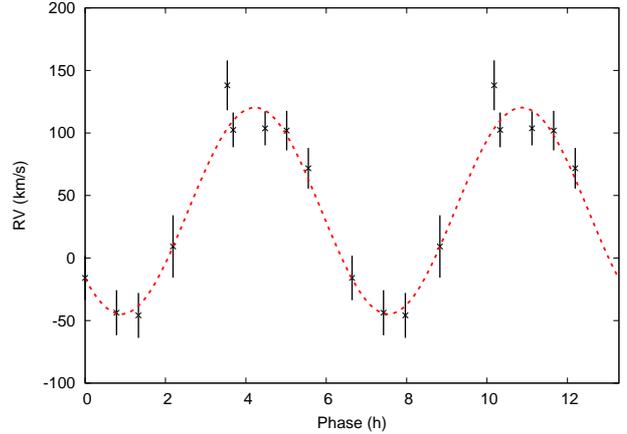}
	\caption{Our orbital fit for J073934.37+172225.5, given the SOAR RV estimates. We obtained $T=6.61\pm0.01$~h, $K=82.6\pm6.8$~km~s$^{-1}$, $RV_0=37.8\pm3.4$~km~s$^{-1}$, and $R^2=0.96$.}    
	\label{J0739}
\end{figure}

We obtained nine spectra in three nights with SOAR for J0739+1722, whose RV variability was already suggested by its SDSS subspectra. The RV estimates from the SOAR spectra give $p = 0.1465$. We obtained a period of $6.64\pm0.03$~h, too short for a main sequence star in the sdA range of parameters \citep{brown2017}, and $K = 82.6\pm6.8$~km~s$^{-1}$. The orbital solution, shown in Fig~\ref{J0739}, has a high $R^2$ of 0.96.

We estimated the mass of the ELM primary to be $0.145\pm0.001~M\subsun$, given the $T\eff =7550\pm12$~K and the $\log~g = 5.06\pm0.05$ estimated from the SOAR combined spectrum. The minimum mass of the companion is $M_2 = 0.10~M\subsun$. For the mean inclination angle for a random stellar sample, $i = 60$, the mass is 0.12~$M\subsun$. Given the orbital parameters, the merging time is smaller than 68~Gyr.


\subsubsection{J084034.83+045357.6} 

\begin{figure}
	\includegraphics[angle=-90,width=\columnwidth]{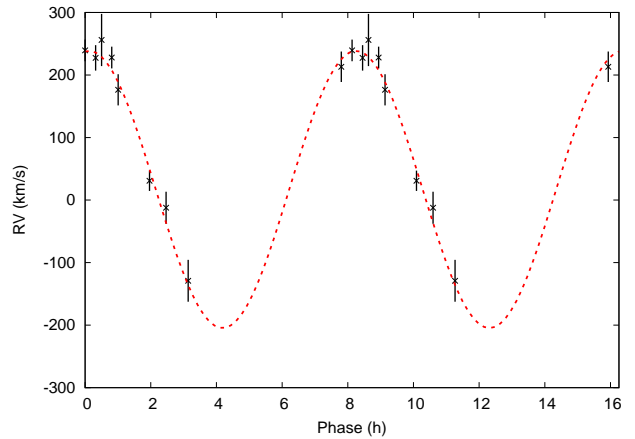}
	\caption{The orbital solution for J084034.83+045357.6, a 0.147~$M\subsun$ pre-ELM with $T\eff \sim 8000$~K. The RV estimates are phase-folded to the 8.13~h period, and show a semi-amplitude of $K=221.6\pm12.8$~km~s$^{-1}$ and $RV_0 = 17\pm13$~km~s$^{-1}$.}    
	\label{J0840}
\end{figure}

J0840+0453 was observed in three nights with SOAR, and we obtained nine spectra. A possible RV variability was first detected in the SDSS subspectra. The Shapiro-Wilk test performed in the SOAR RV data confirmed the variability. The best orbital solution (Fig.~\ref{J0840}) gives $R^2=0.98$, with a semi-amplitude of $221.6\pm12.8$~km~s$^{-1}$ and a period of $8.13\pm0.01$~h.

Our fit to the SOAR spectra of J0840+0453 gives $T\eff = 7890\pm32$~K and $\log~g = 5.07\pm0.09$, implying an ELM mass of $M = 0.147\pm0.002~M\subsun$. The secondary mass is $M_2 > 0.59~M\subsun$, hence it is probably a canonical mass white dwarf. The merging time due to gravitational wave radiation is $\leq 28$~Gyr.


\subsubsection{J134336.44+082639.4} 
\label{2903}

\begin{figure}
	\includegraphics[angle=-90,width=\columnwidth]{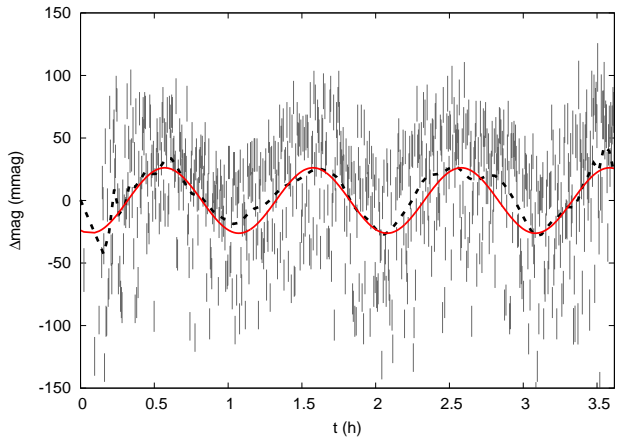}
	\caption{The light curve for SDSS~J134336.44+082639.4, obtained at OPD. There is a lot of spread in the data due to the variation of the seeing throughout the night. The dashed black line shows the smoothed data. The red line shows the best fit obtained with {\sc Period}04, with a period of $3618\pm55$~s and amplitude of $26.1\pm2.4$~mmag.}
	\label{J1343_phot}
\end{figure}

\begin{figure}
	\includegraphics[angle=-90,width=\columnwidth]{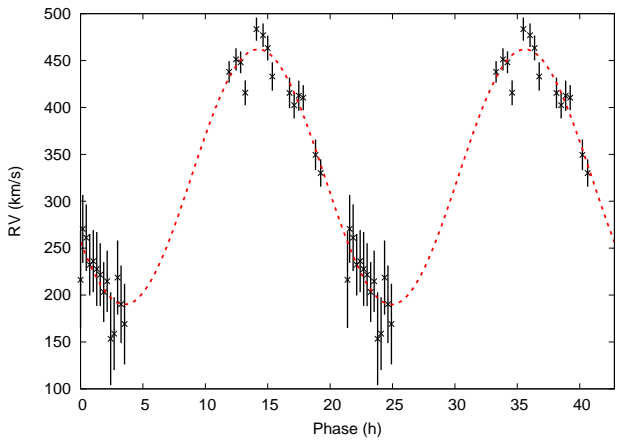}
	\caption{Orbital solution for the photometric variable star SDSS~J134336.44+082639.4, with $T = 21.39\pm0.01$~h, $K = 136.2\pm7.0$~km~s$^{-1}$, $RV_0 = 326.0\pm7.2$~km~s$^{-1}$, and $R^2 = 0.95$.}
	\label{J1343}
\end{figure}

J1343+0826 was found to be photometrically variable in our observations carried out with OPD (see Fig.~\ref{J1343_phot}), and most likely ELM by \citet{pelisoli2018}. We found a photometric period of about one hour, with an amplitude of $26.2\pm2.3$~mmag. This is consistent with the predicted values of \citet{corsico2016}. Unfortunately, this is the only detected period, and therefore we cannot obtain an asteroseismological fit to this object. Spectroscopic follow-up was obtained over five nights at SOAR; twenty-eight spectra were obtained. The derived velocities give $p=0.004$, indicating variability with a high confidence level ($> 99$ per cent). The dominant period was $\sim 24$~h, a probable alias given that four of the observed nights consisted of two sets of consecutive nights.  A similar $R^2$ (only 0.5 per cent smaller) is obtained with $T = 21.39\pm0.01$~h, which is the period we adopt for the orbital solution shown in Fig.~\ref{J1343}. The derived semi-amplitude is $136.2\pm7.0$~km~s$^{-1}$. The systemic velocity is remarkably high, $RV_0 = 326.0\pm7.2$~km~s$^{-1}$. If the $24.7$~h period is the true one, the amplitude would be $175.5\pm6.8$~~km~s$^{-1}$, and $RV_0 = 323.0\pm3.8$~km~s$^{-1}$.

The $\log~g$ we estimated from its SOAR spectrum is among the highest in our sample ($5.97\pm0.03$), and the effective temperature is nonetheless quite low ($T\eff = 8120\pm10$~K), making it a very interesting addition to the known population of pulsating ELMs. We estimate its mass to be $M = 0.153\pm0.001~M\subsun$, while the companion has $M_2 > 0.43~M\subsun$. The objects will merge in less than 449~Gyr.


\subsubsection{J142421.30-021425.4} 

\begin{figure}
	\includegraphics[angle=-90,width=\columnwidth]{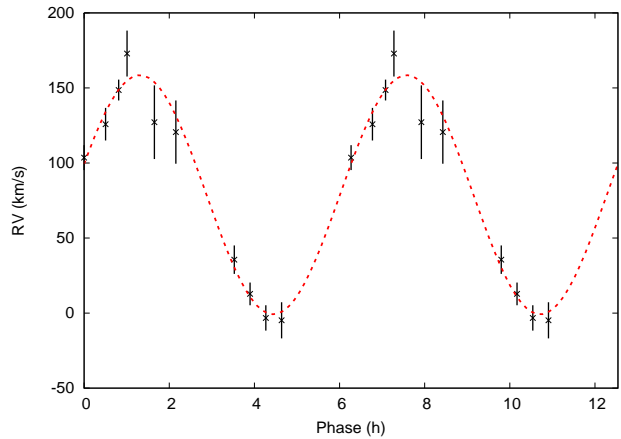}
	\caption{The orbital solution obtained for SDSS~J142421.30-021425.4, whose RV estimates indicated variability at the 85 per cent confidence level.}    
	\label{J1424}
\end{figure}

We followed-up on J1424-0214 given the RV variability suggested by its SDSS subspectra. We observed it in three nights at SOAR, obtaining ten spectra. We obtained a period of $6.3\pm0.4$~h, with a semi-amplitude of $79.7\pm21.8$~km~s$^{-1}$. The orbital solution, shown in Fig.~\ref{J1424}, has $R^2=0.988$.

We derived $T\eff = 9300\pm11$~K and $\log~g =  5.13\pm0.03$ from the SOAR combined spectrum assuming one tenth of the solar metallicity, obtaining $M = 0.1558\pm0.0008~M\subsun$ from the evolutionary models of \cite{althaus2013}. Slightly smaller atmospheric parameters are obtained from the SDSS spectrum, $T\eff = 9090\pm24$~K and $\log~g = 4.53\pm0.04$, resulting on $M = 0.170\pm0.002~M\subsun$. Given the estimated period and semi-amplitude, and assuming the parameters derived from the SOAR spectrum are correct, the companion has $M_2 < 0.09~M\subsun$ ($M_2 = 0.12~M\subsun$ for $q = 60^{\circ}$) and the objects will merge in less than 57~Gyr.


\subsubsection{J205120.67+014554.4} 

\begin{figure}
	\includegraphics[angle=-90,width=\columnwidth]{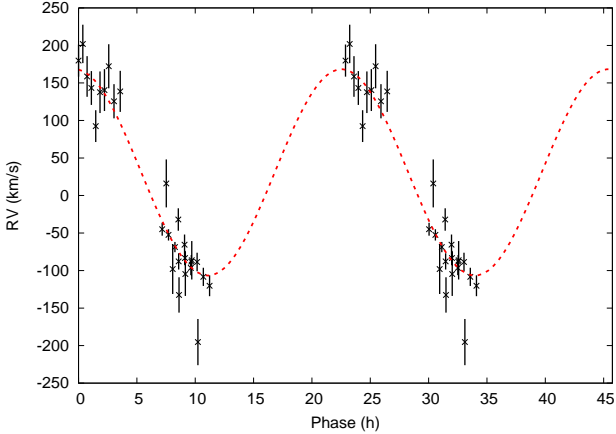}
	\caption{Estimated RVs and orbital solution for SDSS~J205120.67+014554.4, that was observed with both SOAR and Gemini. The orbital fit gives $R^2=0.93$.}    
	\label{J2051}
\end{figure}

J2051+0145 would be at a distance of $\sim 9$~kpc if it had a main sequence radius; however, the estimated proper motion of $9.0\pm2.4$~mas~yr$^{-1}$ suggests a smaller distance, compatible with a (pre-)ELM radius. Observing it for five nights at SOAR, and three nights at Gemini South, we obtained twenty-eight spectra. After obtaining a $p$-value of only 0.002 (confidence level $> 99$ per cent), we estimated the period to be $22.9\pm0.2$~h. The semi-amplitude of the orbital solution is $137.5\pm14.0$~km~s$^{-1}$, and the systemic velocity is $RV_0 = 30.9\pm14.1$~km~s$^{-1}$. The orbital solution is shown in Fig.~\ref{J2051}.

The estimates of $T\eff$ and $\log~g$ using the combined SOAR spectrum, $7810\pm13$~K and $5.00\pm0.05$, give $M = 0.148\pm0.001~M\subsun$. The minimal mass of the secondary is $M_2 = 0.45~M\subsun$, and the merging time due to the emission of gravitational waves is shorter than 533~Gyr.


\subsubsection{J092056.09+013114.8 -- An eclipsing binary} 

\begin{figure}
	\includegraphics[angle=-90,width=\columnwidth]{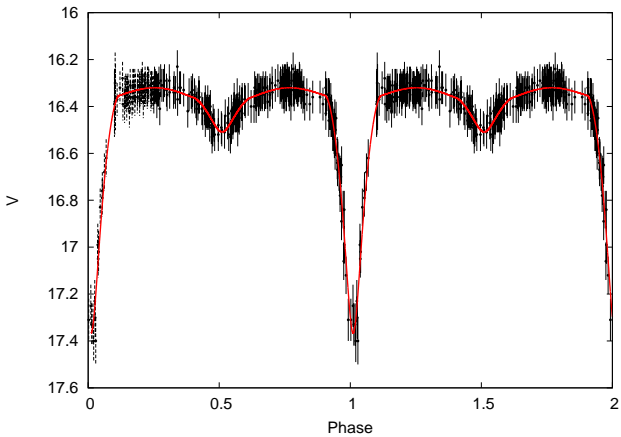}
	\caption{Light curve of J0920+0131, phase-folded to the $15.7$~h period. The best fit to the light curve, calculated with {\sc jktebop}, is shown as a red line.}
	\label{eclips}
\end{figure}

\begin{figure}
	\includegraphics[angle=-90,width=\columnwidth]{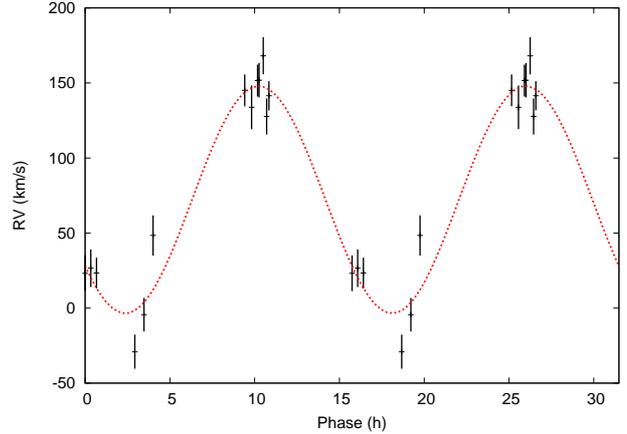}
	\caption{RV data and the obtained orbital solution for SDSS~J092056.09+013114.8. The RV data are phase-folded to the photometric $15.7$~h period.}
	\label{J0920}
\end{figure}

J0920+0131 is an eclipsing binary identified independently by \citet{palaversa2013} and \citet{drake2014}. Using data from the the Catalina Real-Time Transient Survey \citep[CRTS, ][]{drake2009}, we estimate the orbital period to be $15.742\pm0.003$~h. The phase-folded light curve is shown in Fig.~\ref{eclips}. We obtained thirteen spectra in five nights at SOAR. Fixing the period to the photometric estimate, we obtain an orbital solution with $R^2 = 0.95$ (see Fig.~\ref{J0920}) and $K=75.7\pm11.5$~km~s$^{-1}$. 

We fit the photometry using {\sc jktebop} \citep{southworth2004}, and obtain an orbital inclination of $82.7\pm0.4$ and $R_2/R_1 = 0.80\pm 0.03$. Given this inclination, we obtain $M_2/M_1 = 0.894$ from the RV fit. Our spectroscopic fit to the spectrum of the primary gives $T\eff = 7480\pm13$~K and $\log~g=4.80\pm0.06$, implying $M_1 = 0.149\pm0.002~M\subsun$ and $R_1 = 0.25~R\subsun$. Therefore the secondary mass seems to be an even lower mass ELM with $M_2 = 0.133\pm0.002~M\subsun$ and $R_2 = 0.20~R\subsun$. The external uncertainty in the radius, given the 0.25~dex uncertainty in $\log~g$, is about $0.7~R\subsun$ (assuming a fixed mass of 0.15~$M\subsun$). Thus the radius of the secondary might be larger, as it would be expected from a lower mass white dwarf given the mass-radius relation. We estimate the secondary to show $T\eff \sim 4000$~K, given the ratio between fluxes estimated from the light curve.


\subsection{Probable (pre-)ELMs}
\label{probELMs}

\subsubsection{J004227.73-010634.9} 

\begin{figure}
	\includegraphics[angle=-90,width=\columnwidth]{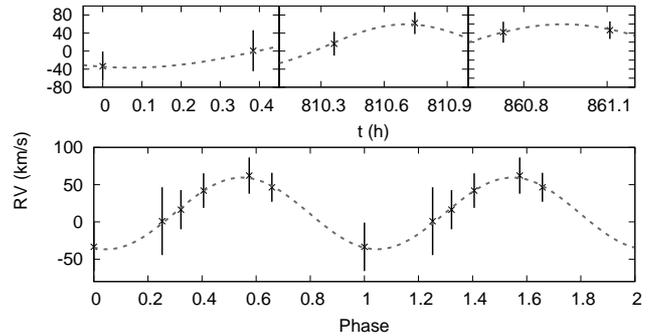}
	\caption{Top panel shows the estimated RVs in the three observed epochs for  SDSS~J004227.73-010634.9. The bottom panel shows a tentative orbital solution, with a 91~min period and $R^2 = 0.993$.}    
	\label{J0042}
\end{figure}

J0042-0106 was followed up because of the $\log g > 5.5$ that we obtained by fitting its SDSS spectrum to solar abundance models. Moreover, assuming the object has a main sequence radius, we obtain a distance of over 15~kpc. It was observed in three nights with Gemini South; two spectra were obtained at each night. Although the observed variations are consistent with Gaussian errors, we obtained an orbital solution with $R^2 = 0.993$ (Fig.~\ref{J0042}). The estimated amplitude is 48~km~s$^{-1}$, and the period is quite low, of only 91~min. Further observations are required to confirm these findings.

The mass of the primary, given the fit to the SDSS spectrum which resulted on $T\eff = 8050\pm24$ and $\log~g=5.51\pm0.08$, is $M = 0.1449\pm0.0003~M\subsun$. Assuming the estimated orbital parameters, we obtain the secondary to show $M_2 > 0.028~M\subsun$ (for an inclination of $60^{\circ}$, $M_2 = 0.033~M\subsun$, and for $15^{\circ}$, $M_2 = 0.22~M\subsun$). Given the short period, the system would merge in less than a Hubble time ($\tau_{\textrm{merge}} < 4.2$~Gyr).


\subsubsection{J011508.65+005346.1} 

\begin{figure}
	\includegraphics[angle=-90,width=\columnwidth]{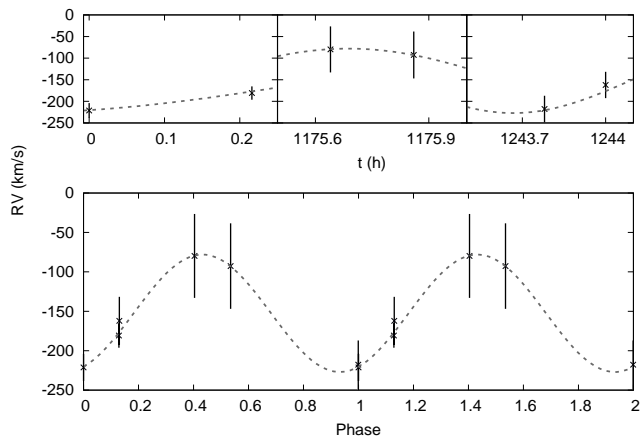}
	\caption{Estimated RVs for SDSS~J011508.65+005346.1 (top), and the best orbital solution (bottom). The average uncertainty was 33~km~s$^{-1}$, much larger than our typical values of 10--15~km~s$^{-1}$, due to the bias issue at GMOS South.}
	\label{J0115}
\end{figure}

J0115+0053 also shows $\log g > 5.5$ in our solar abundance fits to its SDSS spectrum, and $d > 15$~kpc when a main sequence radius is assumed. Moreover, we found it to be more likely ELM in \citet{pelisoli2018}, given its space motion, colours, and physical parameters. We obtained six spectra at three nights with Gemini. The obtained RVs show real variability only at a 70~per cent confidence level, but we obtained $R^2 = 0.97$ for the best orbital solution, with $T = 100$~min and $K = 74$~km~s$^{-1}$ (see Fig.~\ref{J0115}). However, all the spectra were affected by the bright columns that appeared at the Gemini South CCD during the end of 2016B, hence the uncertainties in the velocities are larger.

Given the $T\eff = 8670\pm24$~K and the $\log~g=5.64\pm0.08$, and the hinted RV variability, we propose J0115+0053 is a probable ELM, but we caution that more data is needed to confirm this identification. The evolutionary models give a primary mass of $0.150\pm0.001~M\subsun$, and, assuming the tentative orbital parameters, the minimal secondary mass is 0.049~$M\subsun$ ($M_2 = 0.058~M\subsun$ for $i = 60^{\circ}$, and $M_2 = 0.54~M\subsun$ for $i = 15^{\circ}$), and the merging time is shorter than 3.2~Gyr.


\subsubsection{J030608.92-001338.9} 

\begin{figure}
	\includegraphics[angle=-90,width=\columnwidth]{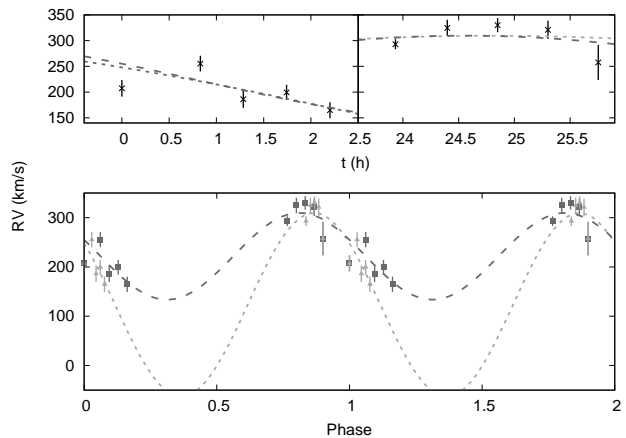}
	\caption{The top panel shows the estimated RVs for SDSS~J030608.92-001338.9 in the two observed nights. Two orbital solutions are shown: short dashed line (light grey) for the $28.1$~h period, and the long dashed line (dark grey) for the $13.1$~h period. In the bottom panel, the RVs are phase folded to these two periods, following the same colour code.}
	\label{J0306}
\end{figure}

We found J0306-0013 to be most likely a (pre-)ELM in \citet{pelisoli2018}. It was classified as a sdA in \citet{dr12cat}. We obtained ten spectra over two nights at SOAR. Fitting the combined spectrum to our $Z = 0.1~Z\subsun$ grid, we obtain $T\eff = 7770\pm10$~K and $\log~g = 5.36\pm0.04$, implying $M=0.1433\pm0.0004~M\subsun$. The Shapiro-Wilk test yields $p<0.3$, suggesting the estimated RVs vary at the 70~per cent confidence level. With only two nights, it is hard to constrain the orbital period. We find two solutions with $R^2$ differing by less than two per cent for $T = 28.6$~h and $T=13.5$~h, the former with $K = 186$~km~s$^{-1}$, and the latter with $K = 88$~km~s$^{-1}$. Both solutions are shown in Fig.~\ref{J0306}. 

For the $28.6$~h period, the secondary has a relatively high minimum mass of $\sim 1.0~M\subsun$. For any inclination above $60^{\circ}$, the companion would have to be a neutron star. The merging time for this period is smaller than 546~Gyr. On the other hand, for the $13.5$~h period, the minimal mass is $M_2 > 0.15~M\subsun$, and the merging time is shorter than 320~Gyr.


\subsubsection{J045515.00-043231.0} 

\begin{figure}
	\includegraphics[angle=-90,width=\columnwidth]{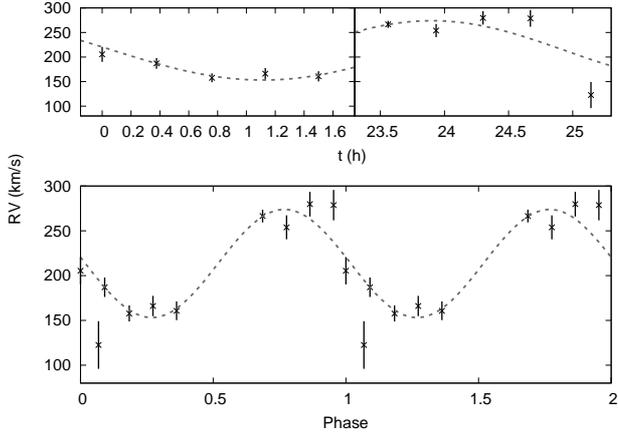}
	\caption{Obtained RVs for SDSS~J045515.00-043231.0 are shown in the top panel. Bottom panel shows the RVs phase-folded to a 4.1~h period.}    
	\label{J0455}
\end{figure}

J0455-0432 was also found to be most likely a (pre-)ELM in \citet{pelisoli2018}. Eleven spectra were obtained over two nights at SOAR. The estimated RVs suggest variability at 65 per cent confidence level. We obtain a dominant period of 4.1~h, but with a high uncertainty of 3.8~h. We estimate $K = 60\pm24$~km~s$^{-1}$. The orbital solution assuming the 4.1~h period, shown in Fig.~\ref{J0455}, gives $R^2=0.89$.

The SOAR spectrum suggests $T\eff = 8250\pm8$~K and $\log~g = 4.15\pm0.03$. These values are consistent with a pre-ELM of $M = 0.180\pm0.001~M\subsun$, in a binary with an object of minimal mass 0.061~$M\subsun$ (0.073~$M\subsun$ for $i = 60^{\circ}$, and 0.69~$M\subsun$ for $i = 15^{\circ}$), which will merge within 25~Gyr.


\subsubsection{J122911.49-003814.4} 

The radial velocity we derived for J1229-0038 from its SDSS spectrum was $472\pm3$~km~s$^{-1}$, consistent with the $465\pm5$~km~s$^{-1}$ given by the SDSS spectral pipeline fit and close to the escape velocity of the Galaxy. Moreover, a main sequence radius would place it at a distance close to 15~kpc, inconsistent with its proper motion of $10.2\pm1.8$~mas~yr$^{-1}$ \citep{tian2017}. Observing it for two nights at SOAR, we obtained five spectra. Although the Shapiro-Wilk test suggests the variability can be explained by Gaussian noise, we find an orbital solution with $R^2 = 0.96$, $T = 3$~h, and $K = 47$~km~s$^{-1}$. We estimate a very high systemic velocity of $510$~km~s$^{-1}$, consistent with the SDSS spectrum, suggesting the semi-amplitude might actually be higher. More data are needed to constrain the orbit of this object

The fit to its SOAR spectrum gives $T\eff=8300\pm21$~K and $\log~g = 5.65\pm0.06$, implying a mass $M = 0.1476\pm0.0009~M\subsun$ in the models of \citet{althaus2013}. Assuming the obtained orbital parameters are correct, the mass of the companion should be higher than $M_2 = 0.035~M\subsun$, or equal to $0.32~M\subsun$ for $i = 15^{\circ}$. The mass for most probable $i = 60^{\circ}$ inclination is $0.042~M\subsun$. The merging time is just above a Hubble time, $\tau_{\textrm{merge}} < 20$~Gyr.


\begin{figure}
	\includegraphics[angle=-90,width=\columnwidth]{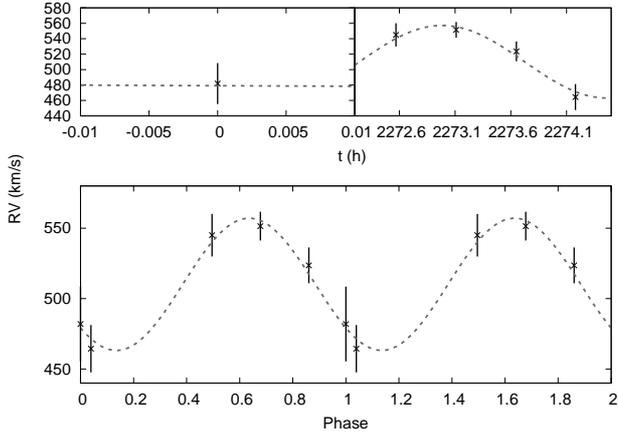}
	\caption{Radial velocities (top) and best orbital solution (bottom) for SDSS~J122911.49-003814.4. The systemic velocity of over 500~km~s$^{-1}$ might indicate that the semi-amplitude is much higher than the derived 47~km~s$^{-1}$.}
	\label{J1229}
\end{figure}

\subsubsection{J233606.13-102551.5} 

\begin{figure}
	\includegraphics[angle=-90,width=\columnwidth]{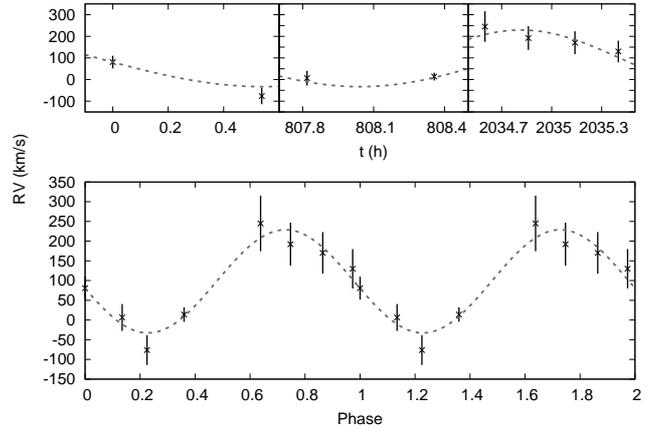}
	\caption{Radial velocity estimates (top) and the best orbital solution (bottom) for SDSS~J233606.13-102551.5.}
	\label{J2336}
\end{figure}

J2336-1025 would be at a distance larger than 26~kpc if it had a main sequence radius. We obtained $\log g = 5.72\pm0.15$ and $T\eff=8330\pm39$~K from its SDSS spectrum using our solar abundance models. This implies $M = 0.149\pm0.003~M\subsun$ and $R = 0.088\pm0.02~R\subsun$. We followed it up for three nights with Gemini South, obtaining six spectra. The RV estimates hint a period of $2.4$~h, even though the variability could also be explained by Gaussian uncertainties. The best orbital solution ($R^2 = 0.92$) gives $K = 131$~km~s$^{-1}$ and is shown in Fig.~\ref{J2336}. With these orbital parameters, we obtain $M_2 > 0.12~M\subsun$ and $\tau_{\textrm{merge}} < 3.73$~Gyr, shorter than a Hubble time.


\subsubsection{J162624.91+162201.5} 

\begin{figure}
	\includegraphics[angle=-90,width=\columnwidth]{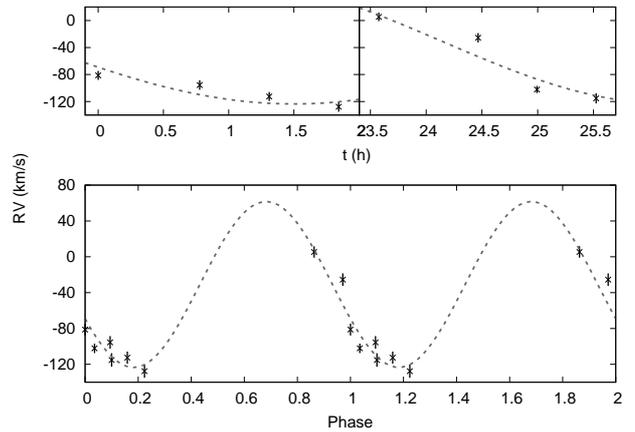}
	\caption{The estimated RVs for SDSS~J162624.91+162201.5 are shown in the top panel, while the bottom panel shows the velocities phase-folded to the 8.2~h period, together with the orbital solution.}
	\label{J1626}
\end{figure}

We obtained in \citet{pelisoli2018} that J1626+1622 is most likely a (pre-)ELM. It was followed-up for two nights at SOAR, when we obtained eight spectra. At a 90 per cent confidence level, the detected RV variability cannot be explained by random uncertainty. We estimated a period of $8.2\pm0.1$~h and $K = 92.6\pm19.3$~km~s$^{-1}$, obtaining an orbital solution with $R^2 = 0.88$, shown in Fig.~\ref{J1626}.

However, J1626+1622 we derived a low $\log~g = 3.83\pm0.03$, with $T\eff= 7460\pm15$~K, from its combined SOAR spectrum. Similar parameters are obtained from the SDSS spectrum. With these parameters, we could only explain it as a pre-ELM in a CNO flash. The time scale of these flashes ranges from $10^5$ to $10^6$ years. The estimated physical parameters are consistent with the flashes of a $M = 0.34~M\subsun$ model. Given the estimated orbital parameters, $M_2 > 0.20~M\subsun$ and $\tau_{\textrm{merge}} < 32$~Gyr.


\subsubsection{J090410.00+034332.9} 

\begin{figure}
	\includegraphics[angle=-90,width=\columnwidth]{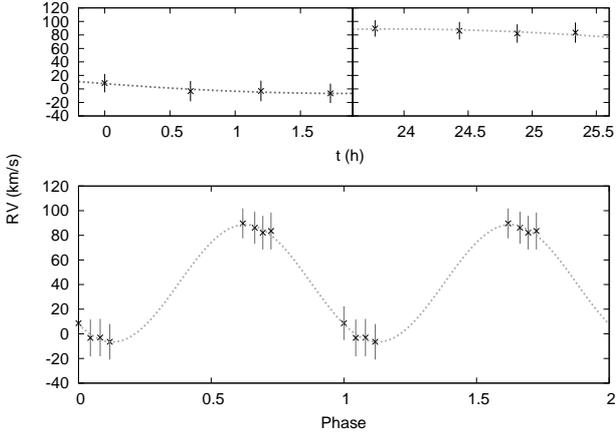}
	\caption{The orbital solution for the SOAR data of SDSS~J090410.00+034332.9, folded to the 14.7~h period (bottom), and the estimated RVs (top).}    
	\label{J0904}
\end{figure}

We obtained eight spectra over two nights at SOAR for J0904+0343, given its RV variability in the SDSS subspectra. The normality test gave $p=0.006$, confirming the variability. We estimated a period of $14.7\pm0.3$~h, with a low semi-amplitude of $47.7\pm2.4$~km~s$^{-1}$, suggesting either a high-orbital inclination, or that the object is a main sequence binary. The orbital solution, shown in Fig.~\ref{J0904} gives $R^2=0.997$.

J0904+0343 SOAR spectrum fits $T\eff = 7680\pm20$ and $\log~g = 4.08\pm0.05$, assuming 0.1~Z$\subsun$. This is compatible with a pre-ELM of mass $0.18\pm0.05~M\subsun$ given the models of \citet{althaus2013}. However, it could also mean that the object is a binary metal poor F star in the halo. The estimated distance given a main sequence radius is 9~kpc, and the proper motion is quite low and uncertain ($3.3\pm 3.1$~mas~yr$^{-1}$). The low detected semi-amplitude results on a low minimal mass of 0.08~$M\subsun$ assuming a pre-ELM primary, given that the orbit would probably not be edge-on. For a $15^{\circ}$ inclination, the mass is about $1.05~M\subsun$, and for $60^{\circ}$, it is 0.092$~M\subsun$. The merging time would be up to 590~Gyr.

We will be able to estimate the distance for this star and others with the parallax to be released by Gaia, and therefore the difference between our preferred solution as a pre-ELM and a 9~kpc main sequence star will be clear.


\subsection{No detected variation}
\label{single}

\begin{figure}
	\includegraphics[angle=-90,width=\columnwidth]{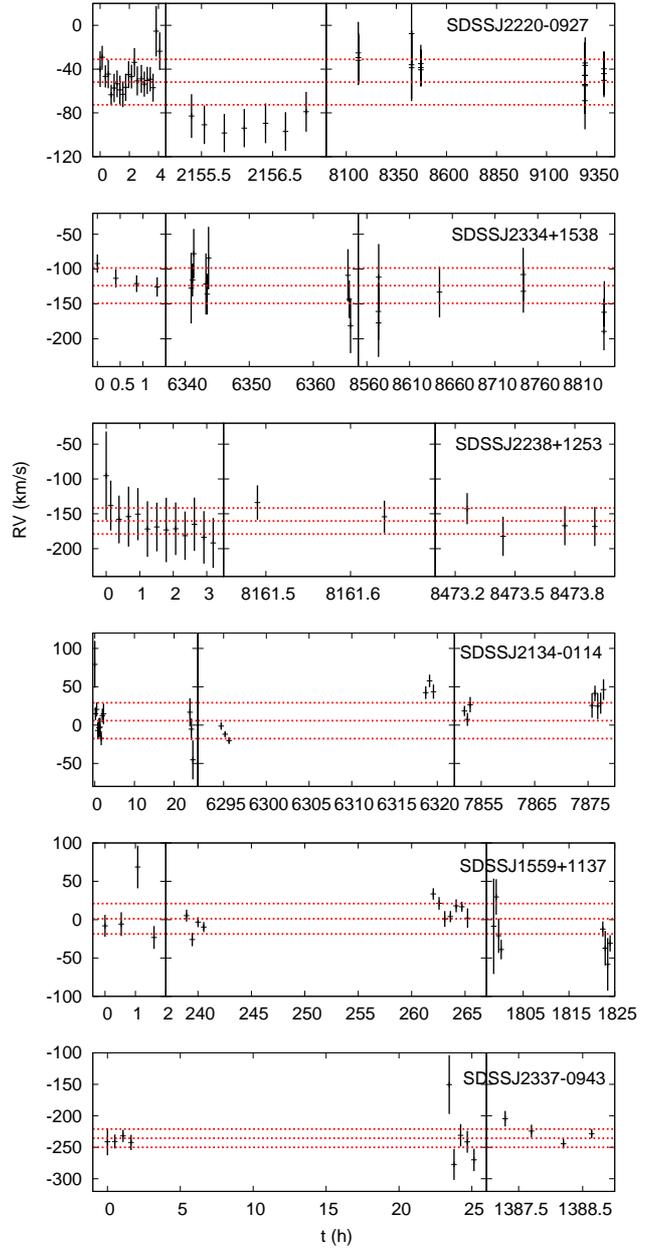}
	\caption{Velocities for the six objects observed in multiple epochs, with no statistically detected variation.}
	\label{singles}
\end{figure}

Fig.~\ref{singles} shows the RV estimates for the objects with no statistically significant RV variations, and no good orbital solutions in the probed ranges of periods, described below. We caution that periods as short as 12~min were observed for the known ELMs \citep{brown2011b}, and theoretical models predict periods of several days \citep{sun2017}.

\begin{figure}
	\includegraphics[angle=-90,width=\columnwidth]{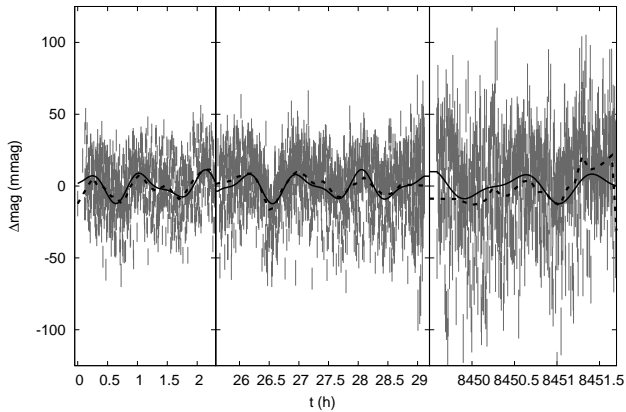}
	\caption{OPD light curve for 222009.74-092709.9. Two periods were found above a detection limit of $4~\langle \textrm{A} \rangle$, where $\langle \textrm{A} \rangle$ is the average amplitude of the Fourier transform. The fit to the light curve given these two periods is shown as continuous line. The dashed line shows the smoothed data.}
	\label{0720lc}
\end{figure}

J222009.74-092709.9 was found to be a photometric variable in OPD data (see Fig.~\ref{0720lc}). Despite its reliable proper motion of $\mu = 9.6\pm1.6$~mas~yr$^{-1}$, we did not find it to be most likely ELM in \citet{pelisoli2018} given its colours. The spectroscopic follow-up revealed no orbital periods in the range $\sim 20$~min to $\sim 200$~days. However, we obtain $\log~g = 6.10\pm0.02$ and $T\eff = 8230\pm6$~K for the SOAR spectra of this object, which not only places it in the region of the known ELMs, but also within the instability strip given by \cite{tremblay2015}, thus justifying the observed photometric variability. We suggest further monitoring of this object should be done to probe shorter and longer orbital periods, as well as further time series photometry to allow an asteroseismological study.

J233403.21+153829.2, J223831.91+125318.3, and J155937.48+113721.9 also have reliable proper motion according to the criteria of \citet{pelisoli2018} \citep[$39.1\pm1.5$, $13.6\pm1.6$, and $7.4\pm1.8$~mas~yr$^{-1}$, respectively, according to][]{tian2017}, yet they were not found to be most likely ELMs. J1559+1137 has not been previously cited in the literature. Our estimated physical parameters are close to the main sequence upper limit, $T\eff=11880\pm41$~K and $\log~g=4.83\pm0.01$. The lack of RV variation and periods in the range $\sim 1~h$ to $\sim 40$~days suggests it is either a short period or ELM, or possibly a halo blue straggler star.

J2343+1538 in particular was suggested to be an extremely-metal poor (EMP) star by \citet{aoki2013}. Their adopted physical parameters based on the SEGUE stellar parameter pipeline \citep[SSPP, ][]{lee2008}, $T\eff = 6500$~K and $\log~g = 4.0$, agree within external uncertainties with the parameters we estimate from SOAR spectra, $T\eff=6710\pm17$~K and $\log~g = 4.25\pm0.05$. No periods were found in the $\sim 24$~min to $\sim 180$~day range, thus the EMP explanation seems likely, although at odds with the high proper motion.

J2238+1253 was photometrically classified as a horizontal branch (HB) star by \citet{xue2008}. However, given our estimated physical parameters from its SOAR spectrum, $T\eff=7870\pm9$~K and $\log~g=5.17\pm0.05$, this classification seems unlikely, considering both the low temperature and high $\log~g$. We find nonetheless no periods in the range  $\sim 20$~min to $\sim 180$~days. The object could either be an EMP with an overestimated $\log~g$, or a single ELM, formed through one of the alternative paths to binary evolution described in the Introduction. {\it Gaia} parallax will allow us to determine its nature.




J213428.63-011409.3 and J233708.62-094307.0 were both followed up considering the high proper motions \mbox{($> 12$~mas~yr$^{-1}$)} displayed in the catalogue of \citet{munn2014}. However, both values were actually unreliable due to close by sources, and the proper motions given in the recent GPS1 catalogue \citep{tian2017} are much smaller and quite uncertain. For J2134-0114, not only no RV variation is found, but the fit to the SOAR spectrum suggests a relatively low $\log~g=3.76\pm0.02$, and $T\eff=12320\pm84$~K, placing it above the zero-age horizontal branch (ZAHB). It could thus be a HB star. We cannot, however, discard the possibility that it is a (pre-)ELM in a CNO flash. J2337-0943, on the other hand, lies below the ZAHB, with $T\eff=8020\pm12$~K and $\log~g=4.59\pm0.06$. Such parameters are consistent with a  $M=0.160\pm0.004~M\subsun$ pre-ELM, but the no detection of orbital periods in the range 1~h to 30~days and the lack of reliable ($> 3\sigma$) proper motion suggest it could be a metal-poor A/F star instead.



\begin{figure}
	\includegraphics[angle=-90,width=\columnwidth]{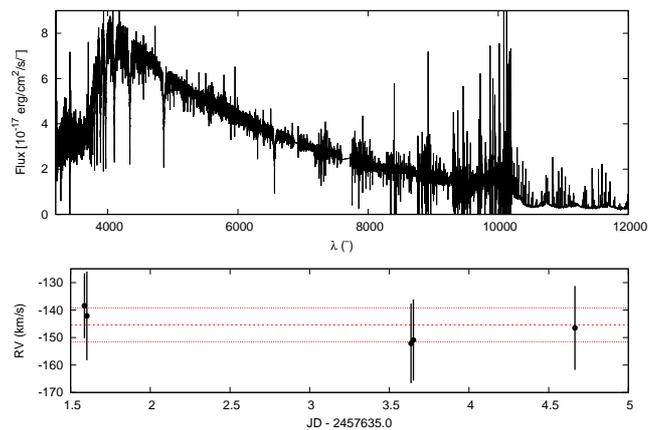}
	\caption{The top panel shows the combined X-shooter spectrum for J024932.84-010708.4. No companion can be identified in the red. The bottom panel shows the RVs obtained for the spectra taken at three different nights. The dashed lines show the weighted mean and the $\pm 1~\sigma$ values.}
	\label{J0249}
\end{figure}

\begin{figure}
	\includegraphics[angle=-90,width=\columnwidth]{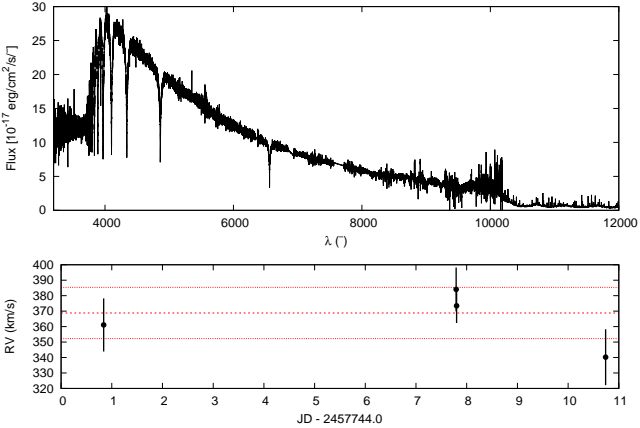}
	\caption{Combined X-shooter spectrum for J101701.89+070806.8 (top), and the RVs obtained from each individual spectrum (bottom). The weighted mean and $\pm 1~\sigma$ values are indicated by the dashed lines.}
	\label{J1017}
\end{figure}

\begin{figure}
	\includegraphics[angle=-90,width=\columnwidth]{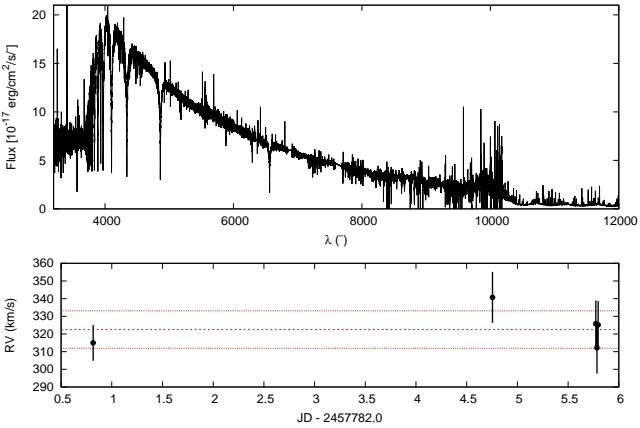}
	\caption{The estimate RVs for J112620.47+090145.5 (bottom), obtained from the individual spectrum taken at three different nights. The Doppler-corrected combined spectrum is shown in the top panel.}
	\label{J1126+0909}
\end{figure}

\begin{figure}
	\includegraphics[angle=-90,width=\columnwidth]{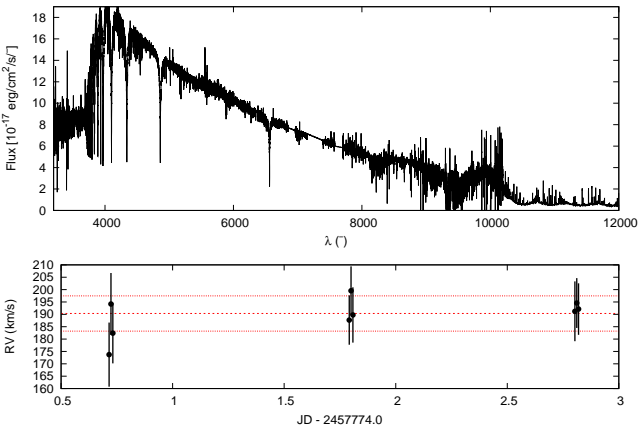}
	\caption{X-shooter spectrum (top) and RV estimates (bottom) for J112616.66-010140.7. The RV estimates agree between the three nights.}
	\label{J1126-0101}
\end{figure}

\begin{figure}
	\includegraphics[angle=-90,width=\columnwidth]{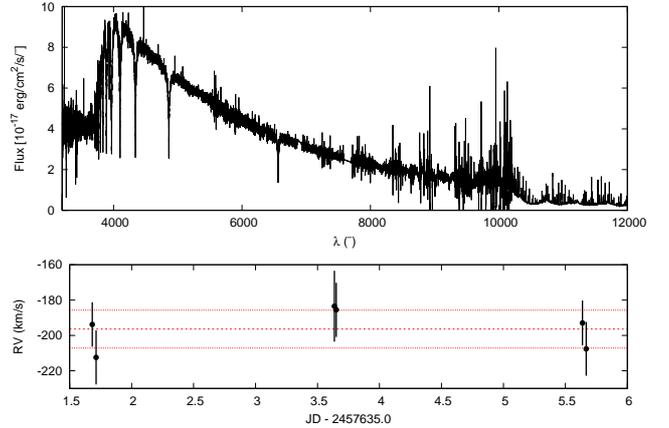}
	\caption{Bottom panel shows the RV estimates for J233343.95-001502.0, and the top panels shows the combined Doppler-corrected spectrum for all epochs.}
	\label{J2333}
\end{figure}

The five objects observed with X-shooter -- J024932.84-010708.4, J101701.89+070806.8, J112620.47+090145.5, J112616.66-010140.7, and J233343.95-001502.0 -- are shown in Figs. \ref{J0249} to \ref{J2333}. Their physical properties as estimated from their X-shooter spectra are given in Table~\ref{xshooter}. Besides no RV variation, it can also be noted that they have no red companions. They all showed $\log g > 5.5$ in our fit to their SDSS spectra assuming solar abundances. Most show $\log g \gtrsim 5.5$ also when $Z = 0.1$~Z$\subsun$ is assumed. Interestingly, the fit to the X-shooter spectra assuming  $Z = 0.1$~Z$\subsun$ suggests a $\log~g$ lower by $\sim 1$~dex. Possible reasons are discussed in Section~\ref{fits}. The obtained parameters and the fact that none shows significant proper motion suggests they could all be metal-poor A/F stars. However, we caution that they are hotter and apparently less metallic than known low-metallicity stars \citep[e.g][]{yong2013}.

\begin{table}
	\centering
	\caption{Physical properties derived for the objects observed with X-shooter, with models assuming $Z = 0.1~Z\subsun$}.
	\label{xshooter}
	\begin{tabular}{cccccc}
		\hline
		\hline
		SDSS~J & $T\eff$ (K) & $\log g$ \\
		\hline
		024932.84-010708.4 & $8219\pm13$ & $4.775\pm0.044$ \\
		101701.89+070806.8 & $8746\pm6$ & $4.331\pm0.020$ \\
		112620.47+090145.5 & $8467\pm7$ & $4.640\pm0.021$ \\
		112616.66-010140.7 & $8073\pm8$ & $4.834\pm0.025$ \\
		233343.95-001502.0 & $8279\pm6$ & $4.410\pm0.016$ \\
		\hline
		\hline
	\end{tabular}
\end{table}

\subsection{Spectral fits}
\label{fits}

\begin{figure*}
	\includegraphics[angle=-90,width=\textwidth]{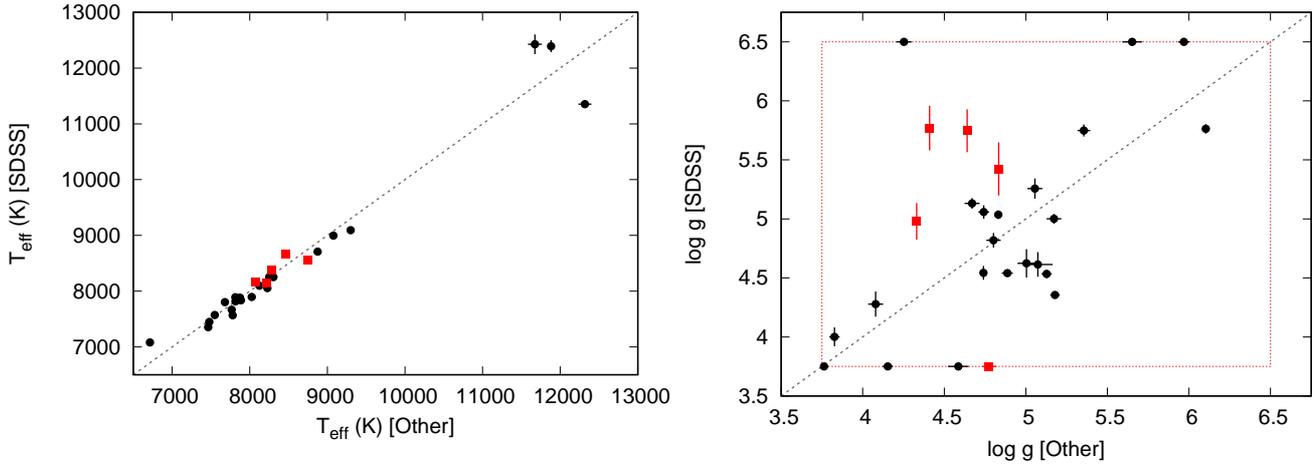}
	\caption{Comparison between the physical parameters obtained by fitting SDSS spectra or SOAR/X-shooter spectra (labelled as other). The physical parameters derived from SOAR spectra are shown as black dots, while parameters derived from X-shooter spectra are shown as red squares. There is a very good agreement, to less than 2~per cent, in $T\eff$ (left). The dashed rectangle in the $\log~g$ panel (right) indicates the border of the model grid. Considering objects within this limits, SOAR and SDSS $\log~g$ show a low average difference of 0.08~dex. The X-shooter spectra suggest a $\log~g$ lower by 0.93~dex, what could be due to the better resolution and larger spectral coverage provided by X-shooter, but could also be explained by statistical fluctuations, as detailed in the text.}
	\label{fit_comp}
\end{figure*}

Fig.~\ref{fit_comp} shows the comparison between our fit to SDSS spectra and to spectra obtained as part of this work, with the same grid of models ($Z = 0.1$~Z$\subsun$). The effective temperature seems to agree remarkably well between spectra obtained with different facilities, with average differences of less than 2~per cent. The $\log~g$, on the other hand, shows a larger spread. Considering the spectra obtained with SOAR, the average difference to SDSS, considering only objects whose fit is not at the border of our grid, is only about 0.08~dex, hence completely consistent with the uncertainties. However, comparing the four X-shooter spectra to SDSS, we obtain a large difference of $-0.93$~dex. The grid of models is the same, so the difference cannot be explained by metallicity as suggested by \citet{brown2017}.

The main differences between X-shooter and SDSS/SOAR spectra are the wavelength coverage and the spectral resolution. At low temperatures, the width of the lines is not very sensitive to $\log~g$, and the $\log~g$ determination depends essentially on flux below 3700~\AA. Spectra obtained with the SDSS spectrograph only cover above $~3800$~\AA. More recent spectra obtained with BOSS extend the coverage down to $3600$~\AA, but usually with low-S/N in this wavelength range. This region is also very sensitive to flux calibration and extinction. Hence $\log~g$ estimates from SDSS spectra might be affected by these uncertainties. We had previously assumed an external 0.25~dex uncertainty \citep{pelisoli2018}, but it appears that it might be even larger, up to 0.50~dex. It is important to caution, however, that these four objects were selected on a very large sample, with tens of thousands of sdAs \citep{pelisoli2018}, hence we should expect to find several objects with errors of $2-3~\sigma$. In short, we cannot assert that this difference between SDSS and X-shooter spectral fits is systematic, as it might result from statistical fluctuations.

 
\begin{figure*}
	\includegraphics[angle=-90,width=\textwidth]{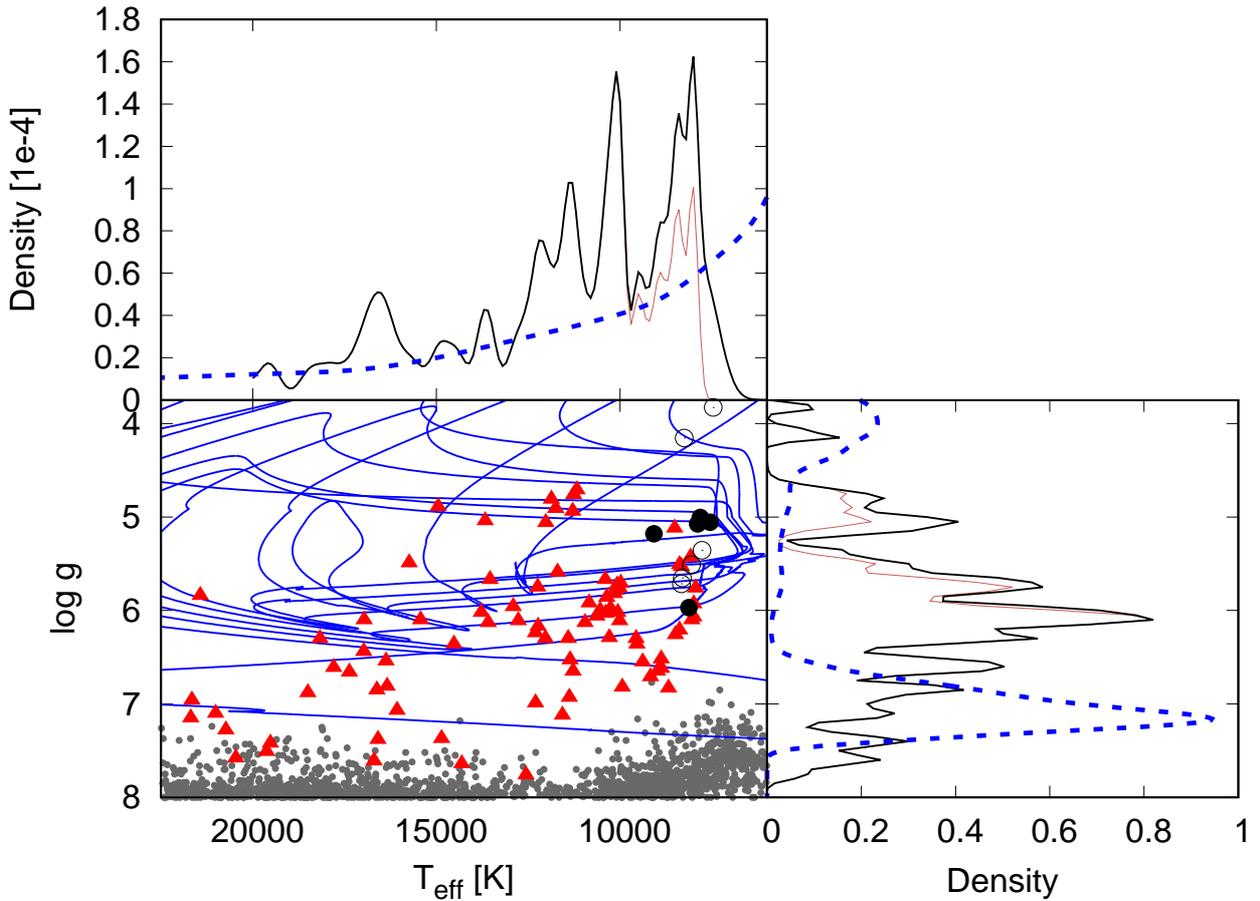}
	\caption{Same as Fig.~\ref{fig1}, adding the new (pre-)ELMs (filled black circles) and probable (pre-)ELMs (open black circles). The distributions with these added objects are shown in black. Especially in $T\eff$, the population seems more similar to the predicted by the models. There are still missing (pre-)ELMs in the lower $\log~g$ end.}    
	\label{fig2}
\end{figure*}

The adopted physical parameters for our new and probable (pre-)ELMs are shown in Table~\ref{elms}, as well as their estimated orbital parameters. Fig.~\ref{fig2} is similar to Fig.~\ref{fig1}, including these new objects. These additions to the known sample of ELMs improve the comparison between model predictions and observed population, by adding objects both to cool and to low-mass ends of the (pre-)ELM space of physical parameters.

\begin{table*}
	\centering
	\caption{Estimated physical and orbital properties for the new and probable (pre-)ELMs, which are separated by a horizontal line. $T\eff$ and $\log~g$ were estimated assuming $Z = 0.1~Z\subsun$. The secondary mass is the lower limit ($i = 90^{\circ}$), and the time for merging is the upper limit. The uncertainties in $P$ and $K$ were calculated with a thousand Monte Carlo simulations in {\sc Period04}.}
	\label{elms}
	\begin{tabular}{cccccccc}
		\hline
		\hline
		SDSS~J & $T\eff$ (K) & $\log~g$ & $M$ (M$\subsun$) & $P_{\textrm{orb}}$ (h) & $K$ (km/s) & $M_2$ (M$\subsun$) & $\tau_{\textrm{merge}}$ (Gyr) \\
		\hline
032914.77+003321.8 & $9077\pm10$ & $5.179\pm0.029$ & $0.1536\pm0.0006$ & $    20.0\pm     0.1$ & $  83\pm 22$ & 0.17 & 765 \\      
073934.37+172225.5 & $7548\pm12$ & $5.056\pm0.046$ & $0.1450\pm0.0011$ & $    6.64\pm    0.03$ & $82.6\pm6.8$ & 0.10 &  68 \\      
084034.83+045357.6 & $7886\pm32$ & $5.074\pm0.091$ & $0.1470\pm0.0022$ & $    8.13\pm    0.01$ & $ 222\pm 13$ & 0.59 &  28 \\      
134336.44+082639.4 & $8123\pm10$ & $5.969\pm0.034$ & $0.1527\pm0.0011$ & $  24.692\pm   0.002$ & $136.2\pm7.0$ & 0.43 & 410 \\     
142421.30-021425.4 & $9299\pm11$ & $5.128\pm0.031$ & $0.1558\pm0.0008$ & $     6.3\pm     0.4$ & $  80\pm 22$ & 0.09 &  57 \\      
205120.67+014554.4 & $7813\pm12$ & $5.004\pm0.055$ & $0.1476\pm0.0014$ & $    22.9\pm     0.2$ & $ 138\pm 14$ & 0.45 & 533 \\      
092056.09+013114.8 & $7478\pm13$ & $4.802\pm0.044$ & $0.1492\pm0.0014$ & $  15.742\pm   0.003$ & $75.7\pm8.1$ & 0.09 &  50 \\      
\hline
004227.73-010634.9 & $8051\pm24$ & $5.510\pm0.081$ & $0.1449\pm0.0003$ & $ 1.52231\pm  0.00002$ & $48.1\pm1.6$ & 0.14 & 4.2 \\      
011508.65+005346.1 & $8673\pm24$ & $5.641\pm0.080$ & $0.1499\pm0.0011$ & $1.678517\pm  0.000009$ & $74.5\pm5.5$ & 0.05 & 3.1 \\      
030608.92-001338.9$^a$ & $7768\pm10$ & $5.356\pm0.039$ & $0.1433\pm0.0004$ & $28.6\pm1.1$ & $186\pm61$ & 1.03 & 546\\ 
                                  &                      &                              &                                 & $13.5\pm2.3$ & $88\pm19$ & 0.15 & 320\\
045515.00-043231.0 & $8251\pm 8$ & $4.154\pm0.031$ & $0.1796\pm0.0014$ & $     4.1\pm     3.8$ & $  60\pm 23$ & 0.06 &  25 \\      
090410.00+034332.9 & $7680\pm20$ & $4.079\pm0.046$ & $0.1810\pm0.0488$ & $    14.7\pm     0.3$ & $47.7\pm2.4$ & 0.08 & 590 \\      
122911.49-003814.4 & $8305\pm21$ & $5.652\pm0.060$ & $0.1477\pm0.0009$ & $    2.96\pm    0.08$ & $  47\pm5.0$ & 0.04 &  20 \\      
162624.91+162201.5 & $7464\pm15$ & $3.827\pm0.032$ & $0.3454\pm0.0127$ & $     8.2\pm     0.1$ & $  93\pm 19$ & 0.20 &  32 \\      
233606.13-102551.5 & $8328\pm39$ & $5.716\pm0.147$ & $0.1487\pm0.0030$ & $ 2.38904\pm  0.0008$ & $ 131\pm 11$ & 0.12 & 3.7 \\     
		\hline
		\hline
	\end{tabular}
\begin{flushleft}
	$^a$ Two distinct periods are possible with the current data. Parameters for both are shown.
\end{flushleft}
\end{table*}

\subsection{Photometry}
\label{phot}

Besides J1343+0826 and J2220-0927 described in Sections \ref{2903} and \ref{single}, we have found five other sdAs to show variability. Pulsation periods and amplitudes for all these seven objects are shown in Table~\ref{periods}. Two objects -- J073958.57+175834.4 and J075519.92+091511.0 -- show large amplitude variations with periods above 2~h. The estimated parameters from SDSS spectra assuming $Z = 0.1~Z\subsun$ are $T\eff = 36\,136\pm67$ and $\log~g = 6.00\pm0.02$ for J0739+1758. For J0755+0915, we could not obtain a good fit with the $Z = 0.1~Z\subsun$ grid --- the $\log~g$ is too close to the lower limit of the grid. For solar metallicity, we obtain and $T\eff = 7470 \pm5$~K and $\log~g = 4.50\pm0.04$. None of these two shows significant proper motion. J0755+0915 could be explained as a metal poor A/F star, but we caution that there is at least one pre-ELM known to show RR~Lyrae-pulsations \citep{pietrzynski2012}, given that during the CNO flashes the pre-ELM can reach the RR~Lyrae instability strip. J0739+1758 was photometrically selected as a possible AM~CVn binary by \citet{carter2013}. However, both the long period and the fact the SDSS spectrum shows no emission lines seem to rule out this possibility. Its temperature places it within the region where subdwarfs stars show pulsations (see Fig.~\ref{strip}), hence it could be a new variable subdwarf star. The spectroscopic fit places the object within the domain of V361 Hya stars \citep[first discovered by][$T\eff > 28\,000$~K]{kilkenny1997}. However, V361 Hya stars usually show $p$-mode pulsations with short periods (100--400~s). A few were found to show also $g$-modes \citep[e.g.][]{Schuh2005}, but the $g$-mode pulsations show low-amplitude, unlike what we found. A photometric fit to this object with fixed $\log~g = 6.0$ suggests a lower temperature of $T\eff = 21\,745\pm280$. Our spectral models do not take into account line blanketing by metals or NLTE effects, which are often important for hot subdwarfs \citep[e.g.][]{nemeth2014}, hence in this case the spectral parameters should be only taken as a rough estimate. The photometric $T\eff$ places the object in the V1093 Her domain \citep[a class first found by][]{green2003}. V1093 Her stars show $g$-mode pulsations of the order of hours, as we have observed. The amplitude we observed is nonetheless higher than for the known V1093 Her stars. Further data is required to better determine the nature of J0739+1758. 

\begin{table*}
	\centering
	\caption{Periods and amplitudes for all objects found to be photometrically variable, as well as $T\eff$ and $\log~g$ derived from the SDSS spectra (SOAR for J134336.44+082639.4 and J222009.74-092709.9), assuming $Z = 0.1~Z\subsun$. For J160040.95+102511.7 and J075519.92+091511.0, a good fit is not obtained with $Z = 0.1~Z\subsun$; we then assumed solar metallicity.}
	\label{periods}
	\begin{tabular}{ccccc}
		\hline
		\hline
		Object & $T\eff$ (K) & $\log~g$ & Period (s) & Amplitude (mmag) \\
		\hline
		J134336.44+082639.4 & $8120\pm10$ & $5.97\pm0.03$ & $26.2\pm2.4$ \\
		\hline
		J222009.74-092709.9 & $8230\pm6$ & $6.10\pm0.02$ & $3591.24\pm0.03$ & $7.9\pm0.7$ \\
		                    &             &              & $2168.8\pm0.2$ & $3.9\pm0.7$ \\
		\hline
		J075738.94+144827.5 & $8180\pm7$ & $4.75\pm0.04$ & $2437\pm15$ & $3.3\pm0.1$ \\
	                        &            &                 & $2986\pm22$ & $2.2\pm0.1$ \\
	                        &            &                 & $2059\pm16$ & $1.7\pm0.1$ \\
	                        &            &                 & $ 802\pm5$	& $0.5\pm0.1$ \\
	    \hline
	    J160040.95+102511.7 & $7816\pm10$ & $4.63\pm0.05 $ & $3849\pm57$ &	$3.1\pm0.2$ \\
	                        &             &              & $2923\pm32$ & $1.4\pm0.2$ \\
	                        &             &              & $2133\pm21$ & $1.0\pm0.2$ \\
	    \hline
	    J201757.29-125615.6 & $8138\pm9$  & $5.14\pm0.05$ & $7171\pm49$ & $9.0\pm0.5$ \\
	                        &             &              & $3011\pm58$	& $4.1\pm0.5$ \\
	    \hline
	    J073958.57+175834.4 & $36\,136\pm67$ & $6.00\pm0.02$ & $>11\,000$ & $>40$ \\
	    \hline
	    J075519.92+091511.0 & $7470\pm5$ & $4.50\pm0.04$ & $>7800$ & $>100$ \\
		\hline
		\hline
	\end{tabular}
\end{table*}

\begin{figure}
	\includegraphics[angle=-90,width=\columnwidth]{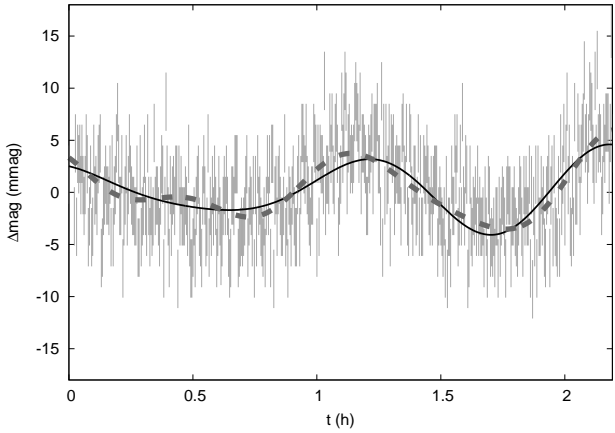}
	\caption{SOAR light curve for 160040.95+102511.7. The grey continuous line is a fit considering the two periods above a 4~$\langle \textrm{A} \rangle$, $3849\pm57$~s with an amplitude of $3.1\pm0.2$~mmag, and $2923\pm32$~s with $1.4\pm0.2$~mmag. The dashed lined adds a third period, which shows an amplitude larger than 3.9~$\langle \textrm{A} \rangle$, $2133\pm21$~s with $1.0\pm0.2$~mmag.}
	\label{2523lc}
\end{figure}

\begin{figure}
	\includegraphics[angle=-90,width=\columnwidth]{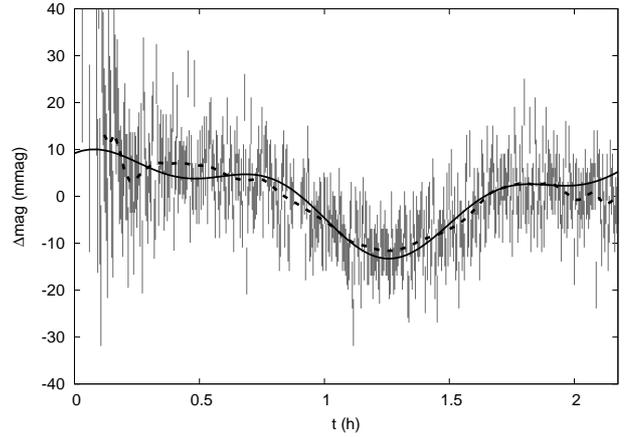}
	\caption{SOAR light curve for and best fit (continuous line) for 201757.29-125615.6. We find two periods, $7171\pm49$~s with and amplitude of $9.0\pm0.5$~mmag and $3011\pm58$~s and amplitude of $4.1\pm0.5$~mmag. Here the dashed line is a smoothing of the data.}
	\label{3137lc}
\end{figure}

\begin{figure}
	\includegraphics[angle=-90,width=\columnwidth]{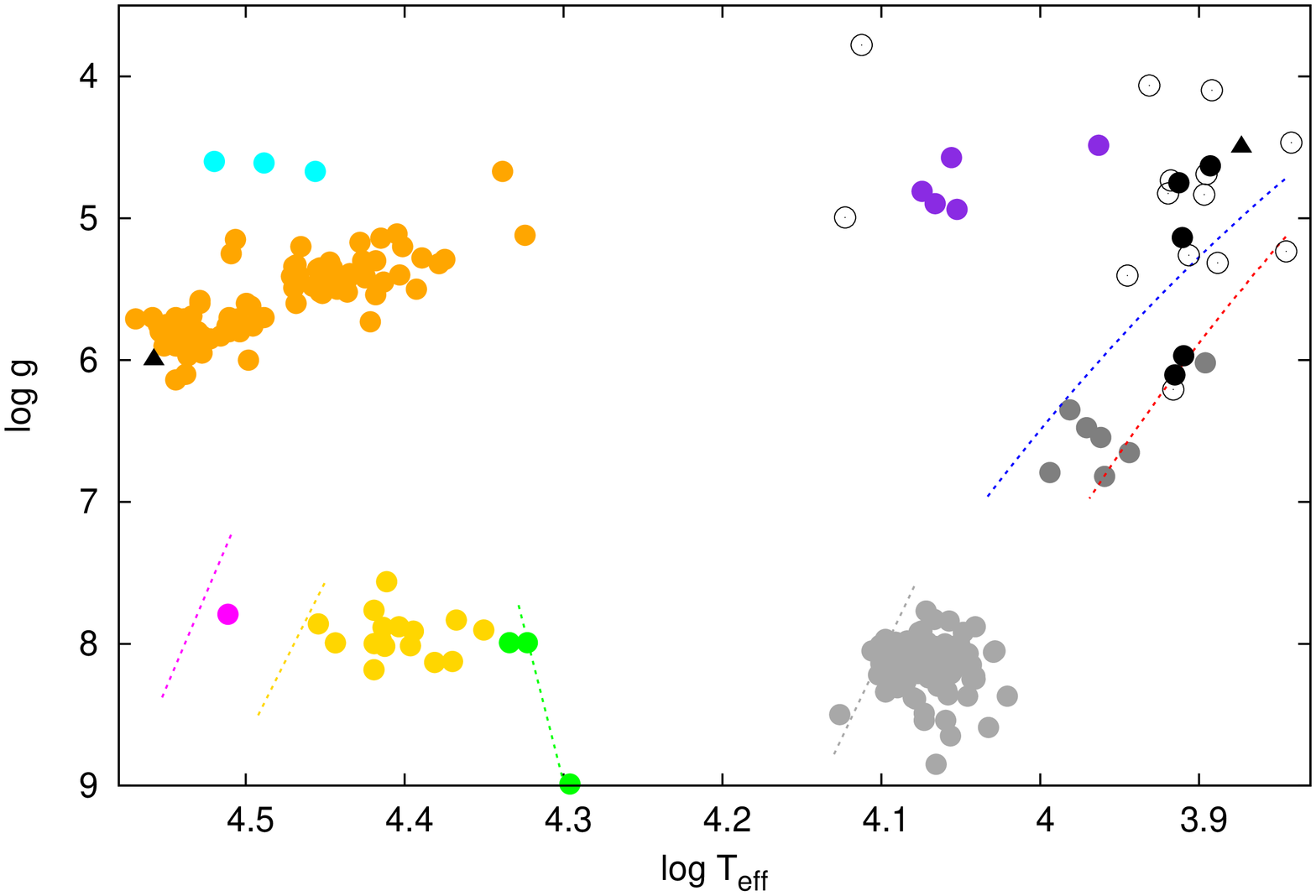}
	\caption{Location of the newly detected variables (filled black symbols) in the $T\eff - \log~g$ diagram. The triangles are the two objects with high amplitude pulsations. Open black circles are objects not observed to vary. The hot (blue line) and cool (red line) edges of the ELM instability strip shown as dotted lines were extracted from \citet{tremblay2015}. Other pulsating white dwarfs are shown as in fig. 1 of \citet{corsico2016}: pre-ELMVs (purple), ELMVs (dark grey), DAVs (light grey, bottom right), DQVs (green, at $\log T\eff \sim 4.3$), DBVs (yellow), and the only known hot DAV (magenta, bottom left). We have also added the sdV stars from \citet{holdsworth2017} (orange), and the newly discovered blue large amplitude pulsators (BLAPs, shown in cyan) from \citet{pietrukowicz2017} as comparison.}
	\label{strip}
\end{figure}

A third object, J075738.94+144827.5 was studied in S\'{a}nchez-Arias et al. (submitted to A\&A) and found to be more likely a $\delta$-Scuti star, given the spacing between periods, a result that should be confirmed by the parallax. Two other new variables, J160040.95+102511.7 and J201757.29-125615.6, are shown in Figs.~\ref{2523lc} and \ref{3137lc}, respectively. The derived physical parameters for the SDSS spectrum of J1600+1025 assuming $Z = 0.1~Z\subsun$ are $T\eff =  7816\pm10$~K and $\log~g = 4.63\pm0.05$, placing it slightly above the instability strip in Fig.~\ref{strip}. Assuming solar metallicity, the parameters are $T\eff = 8050\pm8$~K and $\log~g = 5.59\pm0.03$, placing it within the instability strip of \cite{tremblay2015}. Its proper motion is $7.1\pm1.2$~mas~yr$^{-1}$. For J2017-1256, the derived physical parameters from the SDSS spectra assuming either solar metallicity or $Z = 0.1~Z\subsun$ place it slightly above the instability strip. The $Z = 0.1~Z\subsun$ parameters are $T\eff = 8138\pm9$~K and $\log~g = 5.14\pm0.05$. The proper motion is smaller than $5$~mas~yr$^{-1}$, with an uncertainty of almost $2$~mas~yr$^{-1}$ \citep{tian2017}. The estimated $\log~g$ of both objects is too high for $\delta$-Scuti stars, which have similar spectral properties to (pre-)ELMs, but show $\log~g < 4.4$ \citep[e.g., ][]{murphy2015}. However, given the uncertainties in $\log~g$ described in Section~\ref{fits}, the $\log~g$ could be lower. We obtain a distances of over 3~kpc assuming a main sequence radius, and $z > 1$~kpc given their relatively high galactic latitude. Unfortunately, the number of periods is insufficient for an asteroseismological analysis, thus conclusions on the nature of these objects require more data. 

We have also observed fourteen other objects for at least 2~h, and integration times shorter than 30~s, and found no pulsations. The observing time and obtained detection limits are shown in Table~\ref{phottab}. They are shown in Fig.~\ref{strip} as not observed to vary, but we caution that this does not mean they are not variables. Beating can cause destructive interference and essentially hide the pulsations for hours \citep{castanheira2007}. Moreover, the objects can show pulsations below the detection limit or outside the probed periods.

\begin{table*}
	\centering
	\caption{Objects not observed to vary. The physical parameters were estimated from the SDSS spectra assuming $Z = 0.1~Z\subsun$. For J233625.92+150259.6, we have assumed solar metallicity because no good fit could be obtained with $Z = 0.1~Z\subsun$.}
	\label{phottab}
	\begin{tabular}{cccccc}
		\hline
		\hline
		SDSS~J & $T\eff$ (K) & $\log g$ & Telescope & Exposure time (h) & $3~\langle \textrm{A} \rangle$\\
		\hline
	092140.37+004347.9 & $7733\pm43$ &  $5.315\pm0.179$ & SOAR & 4.0 & 5.0 \\ 
	143333.45+041000.8 & $8814\pm39$   & $5.404\pm0.157$ & SOAR & 1.8 & 7.0 \\ 
	233625.92+150259.6 & $8246\pm 20$ & $6.207\pm0.061$ & OPD & 2.7 & 10.0 \\ 
	112058.97+042012.3 & $13840\pm265$ & $5.147\pm0.065$ & SOAR & 2.4 & 9.0 \\ 
	204038.41-010215.7 & $ 7886 \pm19$ &  $4.833\pm0.096$ & SOAR & 1.6 & 40.0 \\ 
	                   &               &                 & OPD  & 2.2 & 6.0 \\
	163625.08+113312.4 & $ 8538\pm25$ &  $4.064\pm0.060$ & OPD & 4.4 & 12.0 \\ 
	110338.46-160617.4 & $ 8275\pm10$  & $4.734\pm0.051$ & OPD & 5.9 & 10.0 \\ 
	140353.33+164208.1 & $ 8062\pm10$ & $5.261\pm0.035$ & OPD & 3.0 & 4.0 \\ 
	165700.89+130759.6 & $8308\pm11$ &  $4.825\pm0.047$ & SOAR & 2.6 & 25.0 \\ 
	                   &               &                 & OPD & 5.3 & 10.0 \\
	075133.48+101809.4 & $12954\pm131$ & $3.779\pm0.034$ & SOAR & 2.2 & 1.5 \\ 
	045001.34-042712.9 & $ 7797\pm35$ & $4.098\pm0.219$ & SOAR & 3.2 & 8.0 \\ 
	104522.80-023735.6 & $6950\pm36$ & $4.467\pm0.086$ & SOAR & 2.1 & 7.0 \\ 
	094144.89+001233.8 & $7859 \pm32$ & $4.689\pm0.087$ & SOAR & 2.9 & 6.0 \\ 
	111041.50+132354.3 & $ 7859 \pm32$ & $4.689\pm0.087$ & SOAR & 3.2 & 6.0 \\ 
		\hline
		\hline
	\end{tabular}
\end{table*}

\section{Discussion}

With these new pre-ELM and ELM discoveries, we add twelve objects to the $T\eff < 9000$~K range. With the three confirmed ELMs in this range given in \citet{elmsurveyVII}, we reach of total of fifteen objects, compared to 75 in the  $T\eff > 9000$~K range \citep[73 confirmed binaries of][plus J032914.77+003321.8 and J142421.30-021425.4, found in this work]{elmsurveyVII}. This raises the fraction of cool ELMs from 4 per cent to 20 per cent, which is consistent with the predictions by evolutionary models, considering the uncertainties behind the residual burning.

All of our 15 discoveries show $\log~g < 6.0$. Combined with the 26 objects of the ELM Survey in this range, there are now 41 objects in the low-mass end of the ELM distribution. There are 50 objects with $\log~g > 6.0$. The fraction is thus close to 1:1; however, the brightness of the lower $\log~g$ objects suggests the fraction could be as high as 100:1 \citep{pelisoli2018}. Thus, as Fig.~\ref{fig2} already suggested, the population of low-mass objects seems to be missing still. As we considered the estimated $\log~g$ as a selection criterion during most of our follow-up, preferring objects with $\log~g > 5.0$ or even $>5.5$, the fact that this population is not unveiled by our work is not surprising, as there are still thousands of sdAs to be observed. With the upcoming data release 2 of the {\it Gaia} missing, finding this population will be a much easier task.

Out of the five observed objects found to be most likely ELM in \citet{pelisoli2018} (J0115+0053, J0306-0013, J1626+1622, J1343+0826, J0455-0432), one was confirmed as an ELM, and the remaining four were found to be probable ELMs. Four of the objects for which we found no RV variability were also studied in \citet{pelisoli2018}, where a higher probability for the MS channel was obtained. It seems that the probability criteria of \citet{pelisoli2018} are a good indication of the nature of the probed sdAs. On the other hand, ten of the objects we followed up spectroscopically were flagged as possible ELMs in the table 1 of \citet{pelisoli2018} given their $\log~g > 5.5$ estimated from SDSS spectra, but only one was confirmed as ELM (SDSSJ1343+0826), and two others were found to be possible ELMs (SDSSJ0042-0106 and SDSSJ0115+0053). This seems to suggest that the $\log~g$ estimate, especially from SDSS spectra, is not a reliable criteria for selecting ELM candidates. This is in line with our findings described in Section~\ref{fits}. Spectral coverage of the $\lambda < 3700$~\AA{} region with good S/N seems to be a requirement for a reliable estimate of the $\log~g$ in this $T\eff$ range. This might also explain the discrepancies found by \citet{brown2017}, but a study with a statistically significant sample is required to confirm this.

None of the newly discovered (pre-)ELMs has a orbital period short enough ($\lesssim 1$~h) to be above the predicted detection limit of the upcoming LISA gravitational wave detector. However, such short periods were not probed by our survey, given that most objects were observed with SOAR, a 4.1~m telescope, and required integration times close to 30~min to achieve $S/N \gtrsim 10$ in the individual spectra. Searching for this shorter periods might be interesting for the objects described Section~\ref{single}, especially J2220-0927, which not only shows $\log~g > 6.0$ in our spectroscopic fit, but also photometric variability with periods in the ELM range.

The objects for which we found no RV variations could alternatively be metal-poor A/F stars in the halo, as already suggested by \citet{brown2017} as a possible explanation for the sdAs. J2343+1538 in particular seems to be indeed an EMP star, as already suggested by \citet{aoki2013}. The five objects observed with X-shooter, which show no significant proper motion, could also be explained as such. J2134-0114 shows parameters consistent with a HB star, but the nature of the other objects remains puzzling. All show $g < 20.0$, therefore they should be included in the DR2 of {\it Gaia}, making it possible to constrain their radii and determine their nature with certainty. Follow up will still be required for the objects found to be ELMs in order to estimate their orbital parameters, given that {\it Gaia} will not be able to resolve binaries with separations below about 20 milliarcsec ($\sim 2$~AU for a distance of 100~pc).

We have also found seven new photometrically variable stars. J1334+0826 was confirmed as a $M = 0.15~M\subsun$ ELM with time-resolved spectroscopy, so its the eight member of the ELMV class, adding to the seven known pulsating ELMs \citep{hermes2012, hermes2013a, hermes2013b, kilic2015, bell2015}. We found no RV variations for J2220-0927, but its estimated $\log~g$ and temperature place it inside the instability strip, so it is possibly the ninth member of the class. Time-resolved spectroscopy with larger telescopes, allowing shorter integration times, should be done to probe shorter orbital periods for this object. Two other objects, J1600+1025 and J2017-1256, are also found to show pulsations, and are within the instability ELMV strip given uncertainties. \citet{bell2017} found three pulsating stars among the objects in the ELM Survey showing no radial velocity variations, suggesting they are related to the sdA population and might be $\delta$-Scuti stars with an overestimated $\log~g$, which might also be the case for these objects. Hence, as we have not obtained time resolved spectroscopy for these two stars, we make no claim about their nature given the uncertainties in the $\log~g$ estimated from SDSS spectra. Recently, \citet{vos2018} found evidence of a pre-ELM in a long period binary ($771\pm3$~days), likely the result of a merger of the inner binary in a hierarchical triple system. This is also a possible explanation for the systems for which we find no RV variation in short timescales. Another of the variables we discovered, J0739 +1758, seems to be a sdBV, given the $T\eff > 20\,000$~K. Finally, J0755+0915 shows high amplitude pulsations similar to RR~Lyrae stars. The estimated physical parameters and the low proper motion \citep[$3.5\pm1.7$~mas~yr$^{-1}$ according to][]{tian2017} are consistent with a halo metal poor F star.

\section{Summary and Conclusions}

We present radial velocity estimates and spectral fits for 26 sdAs. We find seven to be new (pre-)ELMs, and further eight others to show most characteristics consistent with (pre-)ELMs, but requiring more data to confirm the detection, hence they are called probable (pre-)ELMs. We perform spectroscopic fits, calculate the best orbital solutions, and provide the physical and orbital parameters for each system (Table~\ref{elms}). With this new detections, the percentage of cool ($T\eff < 9000$~K) ELMs is raised from 4 to 20 per cent, which is consistent with the predictions of the evolutionary models. Nonetheless there is still a missing population of (pre-)ELMs in the low-mass end, which should be unveiled by {\it Gaia} DR2. For eleven objects, we find no RV variations, ruling out periods larger than $\sim 1$~h and shorter than $\sim 200$~days for most of them. The high rate of identified binaries in the probed sdAs ($\sim 58$~per cent) suggests that this evolutionary channel indeed plays an important role in explaining the population, as already suggested in \citet{pelisoli2018}.

We have also found the eight member of the ELMV class, J1334+0826. A possible ninth member was also identified, J2220-0927, but its binarity was not confirmed by follow-up time resolved spectroscopy. We found two other sdAs within the ELM instability strip to show pulsations; however, other objects observed within the instability strip have shown no variability, suggesting the strip might not be pure. This should be further investigated when better estimates on the physical parameters of this objects are available, given the possible uncertainty in SDSS spectra that we have identified, due to the lack of good spectroscopic coverage below 3700~\AA.

\section*{Acknowledgements}

IP, SOK, ADR, and LF acknowledge support from CNPq-Brazil. DK received support from programme Science without Borders, MCIT/MEC-Brazil. IP was also supported by Capes-Brazil under grant 88881.134990/2016-01 and would like to thank Bruno C. Quint for the assistance in observing runs with SOAR.

Based on observations obtained at Observat\'{o}rio do Pico dos Dias / LNA, at the Southern Astrophysical Research (SOAR) telescope, which is a joint project of the Minist\'{e}rio da Ci\^{e}ncia, Tecnologia, Inova\c{c}\~{a}o e Comunica\c{c}\~{o}es (MCTIC) do Brasil, the U.S. National Optical Astronomy Observatory (NOAO), the University of North Carolina at Chapel Hill (UNC), and Michigan State University (MSU), and at the Gemini Observatory and processed using the Gemini IRAF package, which is operated by the Association of Universities for Research in Astronomy, Inc., under a cooperative agreement with the NSF on behalf of the Gemini partnership: the National Science Foundation (United States), the National Research Council (Canada), CONICYT (Chile), Ministerio de Ciencia, Tecnolog\'{i}a e Innovaci\'{o}n Productiva (Argentina), and Minist\'{e}rio da Ci\^{e}ncia, Tecnologia, Inova\c{c}\~{a}os e Comunica\c{c}\~{a}oes (Brasil). 




\bibliographystyle{mnras}
\bibliography{sdAs} 




\appendix

\section{Radial Velocity Data}

\begin{table}
	\centering
	\caption{Radial velocity data for the targets in Sections~\ref{newELMs}, \ref{probELMs}, and \ref{single}. The full table is available in the online version of the paper.}
	\label{RVs}
	\begin{tabular}{cccc}
		\hline
		\hline
		Object & BJD & $RV$ (km/s) & $\sigma_{RV}$ (km/s)\\
		\hline
		J032914.77+003321.8 & 2457643.8079 &  247.306 & 58.121 \\
		& 2457643.8153 &  280.711 & 19.405 \\
		& 2457643.8273 &  264.793 & 23.636 \\
		& 2457643.8400 &  260.744 & 21.986 \\
		& 2457643.8521 &  268.121 & 37.191 \\
		& 2457643.8639 &  203.435 & 29.317 \\
		& 2457644.8113 &  161.653 & 11.512 \\
		& 2457644.8232 &  118.973 & 14.990 \\
		& 2457644.8351 &  139.621 & 14.224 \\
		& 2457644.8471 &  111.585 & 15.353 \\
		\hline
		\hline
	\end{tabular}
\end{table}



\bsp	
\label{lastpage}
\end{document}